\documentclass[12pt,twoside]{amsart}
\usepackage{amssymb}
\usepackage[english]{babel}
\usepackage[utf8]{inputenc}
\usepackage{algorithm}
\usepackage[noend]{algpseudocode}
\usepackage{natbib}
\usepackage{comment} 
\usepackage{tikz}
\usetikzlibrary{calc}
\usepackage{kantlipsum} 

\usepackage{lineno}
\usepackage{lipsum}
\def\RR{\mathbb{R}}    
\def\PP{\mathbb{P}}    
\def\EE{\mathbb{E}}    
\def\P{{\mathcal P}}   
\def\R{{\mathcal R}}   
\def\G{{\mathcal G}}   
\def\R{\mathcal{R}}   

\def\proof{\noindent{\em Proof.}~}
\def\eproof{\mbox{\ }\hfill$\square$}
\newtheorem{theorem}{Theorem}
\newtheorem{lemma}{Lemma}
\newtheorem{proposition}{Proposition}
\newtheorem{corollary}{Corollary}
\newtheorem{assumption}{Assumption}
\newtheorem{definition}{Definition}
\newtheorem{example}[theorem]{\textsc{Example}}
\date{}

\title{Learning in Random Utility Models Via\\  Online Decision Problems }

\author{Emerson Melo}

\date{\today}

\calclayout
\calclayout
\thanks{Department of Economics,
	Indiana University, Bloomington, IN 47408, USA. Email: {\tt emelo@iu.edu}. I am very grateful to Bob Becker, Austin Knies, Jorge Lorca, and Pablo Pincheira  for their  valuable comments and suggestions that have greatly improved the paper. }
\begin{document}

\maketitle

\pagestyle{myheadings} \thispagestyle{plain} \markboth{ }{ }
\begin{abstract} 
	This paper studies the Random Utility Model (RUM) in a repeated stochastic choice situation, in which the decision maker is imperfectly informed about the payoffs of each available alternative. We develop a gradient-based learning algorithm by embedding the RUM into an online decision problem. We show that a large class of RUMs are Hannan consistent (\citet{Hahn1957}); that is, the average difference between the expected payoffs generated by a RUM and that of the best-fixed policy in hindsight goes to zero as the number of periods increase. In addition, we show that our gradient-based algorithm is equivalent to the Follow the Regularized Leader (FTRL) algorithm, which is widely used in the machine learning literature to model learning in repeated stochastic choice problems. Thus, we provide an economically grounded optimization framework to the FTRL algorithm. Finally, we apply our framework to study recency bias, no-regret learning in normal form games, and prediction markets.
	
	\end{abstract}
%
\vspace{3ex}

\thispagestyle{empty}

\newcommand{\spacing}[1]{\renewcommand{\baselinestretch}{#1}\large\normalsize}
\textwidth      5.95in \textheight 600pt
\spacing{1.0}
 \newpage
\section{Introduction}\label{s1}
The random utility model (RUM) introduced by \citet{Marschak1959}, \citet{BlockMarschak1959}, and \citet{BeckerGordonDegroot1963} has become the standard approach to model stochastic choice problems.\footnote{ The seminal paper by \citet{Tversky1969}  reports early evidence on stochastic choice behavior. Recently,  \citet{AgranovOrtoleva2016} provided experimental evidence supporting that decision-makers exhibit stochastic choice behavior in repeated choice situations. } The seminal work by \citet{McFadden1978, mcffadens1, mcf1}' takes the RUM approach to a whole new level by making this theory empirically tractable. In particular, he provides an economic foundation and econometric framework which connects observables to stochastic choice behavior. This latter feature makes the RUM suitable to deal with complex choice environments and welfare analysis (\citet{McFadden2001} and \citet{train_2009}).
\smallskip 

 In a RUM, a  decision maker (DM) faces a discrete choice set of alternatives in which each option is associated  with a \emph{random} utility. Then the DM chooses a particular option with a probability equal to the event that such alternative yields the highest utility among all available alternatives. Most of the applied literature models the random utility associated with each alternative as the sum of an \emph{observable} and \emph{deterministic} component and a \emph{random preference shock}. Under this additive specification, different distributional assumptions on the random preference term will generate different stochastic choice rules. Thus, all the effort is to provide conditions on the distribution of the random preference shock such that the choice probabilities are consistent with the random utility maximization hypothesis (\citet{mcf1}). 
From the description above,  we remark that a RUM relies on two fundamental assumptions in addition to the distributional requirements. First, the RUM represents a \emph{static} choice situation, ruling out dynamic environments where the DM may face a repeated stochastic choice problem. Second, the  RUM assumes that the utilities (deterministic plus random components) are known to the DM. However,  it is not difficult to find situations where informational frictions prevent the DM  from learning the utilities associated to the different alternatives. In particular,  the DM may be imperfectly informed about the actual value of the deterministic component associated with each option. These informational frictions may be caused by attentional limits, personal inclinations or biases, or just by the inherent complexity of the alternatives presented to the DM.\footnote{The recent literature in stochastic choice incorporates informational frictions by explicitly modeling the sources 
	of information. For instance, the papers by \citet{Matejka2015}, \citet{CaplinDean2015}, \citet{Caplinetal2019}, \citet{Fosgerauetal2019a},  and \citet{Natenzon2019} study static stochastic choice problems with costly information frictions by incorporating the mechanism by which the DM acquires information and learns about the utilities associated to the alternatives. \citet{WEBB2019} derives a RUM  using a bounded accumulation model, which can capture the dynamic of evidence accumulation. Webber's framework provides an alternative mechanism to information acquisition in stochastic choice models. In a recent paper, \citet{cerreiavioglio2021multinomial} axiomatize the multinomial logit (MNL)  model,  where stochastic choice behavior  is caused by time-constrained information processing. They provide a neural and behavioral foundation for the  MNL in environments where the  DM faces a pressing deadline that affects her choices.}
\smallskip

This paper studies a RUM which relaxes the assumptions of static choice and perfect information. In doing so, we embed the RUM  into an online decision problem (ODP) in which the  DM must choose a probability distribution over a set of discrete actions at each point in time. At each period, there is a random utility vector describing the utilities associated with each option. However,  and different from the traditional RUM, the deterministic component in the random utility vector is unknown to the DM at the moment of making a choice. The realization of this random vector depends on a probability distribution unknown to the DM. In this environment, when making a decision (selecting a probability distribution),  the DM uses the accumulated information up to the previous period, which is subject to random preference shocks. Thus, at each point of time, the DM's probability distribution can be considered generated by a RUM in which she \emph{imperfectly} estimates the performance (utilities) associated to each alternative using the accumulated information until the previous period. To make the connection between both approaches explicit, we denote the resulting model as RUM-ODP.

\smallskip

In the RUM-ODP   model, the DM wants to choose a sequence of probability distributions to suffer as little \emph{regret} as possible, where regret is defined as the difference between the DM's cumulative expected payoffs and that of the best-fixed action in \emph{hindsight} (\citet{Bell1982}, \citet{LoomesSugden1982}, and \citet{FISHBURN198231}). More importantly, when the DM can find a sequence of probability vectors such that the \emph{average} regret associated to it becomes arbitrarily small as  $T$ grows, we say that such a sequence is  \emph{Hannan consistent} (\citet{Hahn1957}). Similarly, when a sequence of probability distributions is Hannan consistent,  we say such a sequence enjoys the \emph{no-regret learning} property (\citet{BianchiLugosi2003} and \citet{Roughgarden2016}).
\smallskip

Examples that fit into the RUM-ODP framework include prediction from expert advice, repeated consumer choice, adversarial learning, and online-shortest path problems, among many others (\citet{Hazan2017}). Furthermore, from a strategic point of view, the RUM-ODP problem can be cast as a structured repeated game between the DM and the environment (nature), where the performance metric (or equilibrium concept) is the regret suffered by the  DM.\footnote{ This strategic interpretation in the ODP literature has been analyzed by \citet{LittlestoneWarmuth1994}, \citet{FreundSchapire1997}, \citet{FreundSchapire1999}, \citet{FosterVohra1999}, and \citet{CesaBianchiLugosi2006} among many others.} 

\smallskip
A fundamental property of the RUM-ODP is the fact that we do not need to specify the DM's priors over the set of possible utility vectors. Similarly, in the RUM-ODP, we do not need to specify the mechanism by which the DM acquires information. This  is in stark contrast with   other models that connect RUM with stochastic choice under informational frictions (e.g. \citet{Matejka2015} and \citet{Fosgerauetal2019a}).
\smallskip

This paper studies the RUM-ODP model from an algorithmic standpoint, making at least five general contributions. First,  we show that the  RUM-ODP model can be studied  in terms of an algorithm which we call the \emph{Social Surplus Algorithm} (SSA). This algorithm provides a simple economic and behavioral foundation to study learning in the RUM-ODP. In doing so, we use the result that, at each point in time, the DM's stochastic choice rule is given by the gradient of the Social Surplus function. This latter relationship  characterizes the entire class of RUMs.\footnote{In a RUM, the Social Surplus function is defined as $\varphi(\bold{v})=\EE_{\epsilon}(\max_{i\in A}\{\bold{v}_i+\epsilon_i\})$, where $A=\{1,\ldots, N\}$ is the choice set,  $\bold{v}_i$ is a deterministic term and $\epsilon_i$ is a random preference shock. \citet{mcf1} and \citet{Rust1994}  show that ${\partial \varphi(\bold{v})\over \partial \bold{v}_i}=\PP(i=\arg\max_{j\in A}\{\bold{v}_j+\epsilon_j\})$ for $j=1,\ldots, N$.}  Then, by exploiting convex analytic results, we show that the SSA is Hannan consistent for a large class of RUMs. A fundamental property of the SSA  is that it allows for environments with arbitrary degrees of correlation (similarity) between the alternatives. Thus, we can study the RUM-ODP in situations where the alternatives in the choice set may be correlated, have complex substitution patterns, have an ordered structure, or be grouped into several different classes (nests). \smallskip

In our second contribution, we focus on a particular instance of the RUM-ODP model. In modeling the random preference shock, we focus on the class of RUMs known as  Generalized Extreme Value (GEV) models. Examples within this class include the multinomial logit (MNL),  the Nested Logit (NL),  the Paired Combinatorial Logit (PCL),  the Generalized Nested Logit (GNL), and the Ordered GEV (OGEV) models. We show that the  SSA is Hannan consistent for all of these models. From an applied perspective, this result is useful in modeling dynamic discrete choice demand and learning in complex environments. To the best of our knowledge, the GEV class has not been studied previously in the context of a model like the RUM-ODP.
\smallskip

Third, we show that the SSA approach is equivalent to the Follow the Regularized Leader (FTRL) algorithm, widely used in machine learning problems involving repeated stochastic choice situations. We show that the SSA and the FTRL  are \emph{dual} to each other in a convex duality sense. The relevance of this equivalence result comes from the fact we can avoid the specification of a regularization function which is fundamental to implementing the FTRL algorithm in applied settings.\footnote{ The FTRL algorithm models the  DM  as solving an explicit strictly concave optimization problem combining  past information (cumulative payoffs) along with a  deterministic regularization term. The idea of regularization comes from the machine learning literature (\citet{Shalev-Shwartz2012}). Intuitively, this idea can be interpreted as hedging against bad future events or as avoiding overfitting the observed data.    Thus, the solution to this regularized optimization problem at each point yields the DM's optimal probability distribution as a function of cumulative payoffs. } In fact, the only regularization function available in closed form is the Shannon entropic term which yields the MNL (cf. \citet[p. 101]{Hazan2017}. More importantly, from an applied standpoint, our equivalence result allows us to provide a behavioral interpretation of the FTRL in terms of a RUM-ODP model. In addition, we show that for a large class of RUMs, the choice probability vector generated by the FTRL algorithm can be written in a general recursive way. This latter result generalizes in a nontrivial way the popular \emph{Exponential Weights Algorithm} (EWA) approach, which heavily relies on the closed-form expression of the MNL model (cf. \citet{Hazan2017} and \citet{Roughgarden2016}). Furthermore, we discuss how to incorporate 
the phenomenon of recency bias, which refers to a situation in which a DM reacts more heavily to recent observations than she does to old ones (\citet{ErevHaruv2016}). In Proposition \ref{OFTRL_Regret},  we show how the RUM-ODP is Hannan consistent under different forms of recency bias. This result provides an economic  foundation to  the  \emph{Optimistic} FTRL algorithm proposed by \citet{RakhlinSridharan2013}.

 \smallskip

In our fourth contribution,  we provide a complete analysis of the NL model in terms of the FTRL algorithm. \citet{McFadden1978}  introduces the NL model as a particular instance of the GEV class. The fundamental property of the NL is that the choice set is divided into a collection of \emph{mutually} exclusive nests. The utilities between alternatives within the same nest are correlated, while the utilities of alternatives in different nests are independent. In this case, we show that the FTRL can be implemented using a new regularization function,  which naturally generalizes the   Shannon Entropic term. From a  behavioral standpoint,  our regularization function captures  substitution patterns \emph{within} and \emph{between} nests.  Furthermore, we show that in the case of the NL model,  the choice probability vector (the gradient of the social surplus function) can be written in an explicit recursive form, highlighting the role of past information and the nesting structure in the DM's learning

 In our last contribution, we focus on two concrete problems. First,  we apply our framework and results to study no-regret learning in normal form games. In particular, we discuss how our approach differs  from the \emph{Potential-Based  dynamics}  framework introduced by \cite{BianchiLugosi2003}, \cite{HART_JET_200126}, and \cite{HART_GEB_2003375}.   Second, use our approach to study predictions markets. A prediction market is a future market in which prices aggregate information and predict future events (\citet{Hanson02logarithmicmarket}). Applications of these markets include electoral markets, science and technology events, sports events, the success of movies, etc. (\citet{WolfersZitzewitz2004}). We show how the RUM-ODP  and the SSA approach are useful for studying this class of markets. In Proposition \ref{SS_equiv_CBM} we exploit the mathematical structure of the social surplus function to connect the SSA with a large class of prediction markets. In economic terms, this result establishes a formal relationship between machine learning, the RUM-ODP  model, and prediction markets. Our results extend in a nontrivial way   the findings in  \citet{ChenWortman2010} and \citet{Abernethyetal2013}.
 \smallskip
 
The rest of the paper is organized as follows. \S \ref{s2} describes the model and studies the SSA. \S \ref{s33} studies the connection between the SSA using the GEV class. \S\ref{s4} analyzes the FTRL algorithm and NL model. In \S \ref{S5_Games} discusses no-regret learning in normal form games. \S \ref{s555} discusses the connection between our results and prediction markets. \S \ref{s5} provides an in-detail discussion of the related literature to this paper. Finally, \S\ref{s6} concludes. Proofs and technical lemmas are gathered in Appendix \ref{Proofs}.
\smallskip

\noindent\textbf{Notation.} Let $\langle\cdot,\cdot\rangle$ denote the inner product between two vectors. For a convex function $f:K\subseteq\RR^n\longrightarrow\RR$, $\partial f(\bold{x})$ denotes the subgradient of $f$ at $\bold{x}.$  Let $\nabla f(\bold{x})$ denote the gradient of a function $f:K\subseteq\RR^N\longrightarrow\RR$ evaluated at point $\bold{x}$.  The $i$th element of $\nabla f(\bold{x})$ is denoted by $\nabla_if(\bold{x})$.  The Bregman divergence  associated to a function $f$ is given  by $D_f(\bold{x}||\bold{y})=f(\bold{y})-f(\bold{x})-\langle\nabla f(\bold{x}),\bold{y}-\bold{x}\rangle$.  The Hessian of  $f$ at point $\bold{x}$ is denoted by $\nabla^2f(\bold{x})$ with entries given by $\nabla^2_{ij}f(\bold{x})$ for $i,j=1,\ldots,N$. Let  $\|\cdot\|$ denote a norm in $\RR^N$ where $\|\cdot\|_*$ is its dual norm.  Let $\bold{A}\in\RR^{N\times N}$ denote a $N$-square matrix. We define the norm $\|\cdot\|_{\infty,1}$ associated to the matrix  $\bold{A}$ as $\|\bold{A}\|_{\infty,1}=\max_{\|\bold{v}\|_1\leq 1} \|\bold{A}\bold{v}\|_1$. Finally, the trace of a matrix $\bold{A}\in\RR^N$ is denoted by $Tr(\bold{A})$.  
\section{Online decision problems and  the Social Surplus Algorithm}\label{s2}

Let $A=\{1,\ldots,N\}$ be a finite set of alternatives. Let $\Delta_N$ denote the $N$-dimensional simplex over the set  $A$. Let $T\geq 2$ denote the (exogenous) number of periods. Let $\bold{u}_t=(\bold{u}_{1t},\ldots\bold{u}_{Nt})$ be a random vector, where $\bold{u}_{it}$ denotes the stochastic payoff associated to option $i\in A$ for $t=1,\ldots, T$. The realizations of the vector $\bold{u}_t$ are determined by the environment (nature), which in principle can be adversarial. We assume that the  vector $\bold{u}_t$ takes values on a compact set  $\mathcal{U}\subseteq \RR^N$. In particular,  we assume that   $\|\bold{u}_t\|_{\infty}\leq u_{\max}\quad\mbox{for all $t$.}$
\smallskip

 In this paper, we study the following ODP. At each period of time $t=1,\ldots, T$,  the DM  chooses  $\bold{x}_t\in \Delta_N$ based on the data received up to the previous period. After committing to  $\bold{x}_t$, the DM observes the realization of the payoff vector  $\bold{u}_t$. Then the DM experiences the expected payoff  $\langle \bold{u}_t,\bold{x}_t\rangle$.  The  DM's goal is to choose a sequence  $\bold{x}_1,\ldots,\bold{x}_T$  that minimizes her \emph{regret} between the total expected payoffs she has incurred and that of the best choice in \emph{hindsight}. Formally, the regret associated with a sequence of choices is defined as follows.
\begin{definition}\label{Regret_Definition} Consider $T$ periods and a sequence of choices $\mathcal{A}=\{\bold{x}_1,\ldots,\bold{x}_T\}$. The regret associated with  the  sequence $\mathcal{A}$ is defined as: 
	\begin{equation}\label{Regret_eq}
	\textsc{R}_{\mathcal{A}}^T=\max _{\bold{x} \in \Delta_N}\left\langle \boldsymbol{\theta}_T, \bold{x}\right\rangle-\sum_{t=1}^{T}\left\langle \bold{u}_{t}, \bold{x}_{t}\right\rangle,
	\end{equation}
	where  $\boldsymbol{\theta}_{T}\triangleq\sum_{t=1}^{T} \bold{u}_{t}$ is the cumulative payoff vector until period $T$. 
\end{definition} 

In the previous definition it is easy to see that  $\textsc{R}_{\mathcal{A}}^T$ can  be interpreted as  the comparison between the sequence  $\bold{x}_1,\ldots,\bold{x}_T$  against the \emph{best choice in hindsight} $\bold{x}^*\in\arg\max \langle\boldsymbol{\theta}_T,\bold{x}\rangle$. Note that   $\bold{x}^*$ is computed in the ideal situation where the DM knows in advance the sequence of payoffs $\bold{u}_1,\ldots,\bold{u}_T$. In particular, in an ODP, the sequence $\mathcal{A}$ describes how the DM learns over time using past information. Thus, $\mathcal{A}$ can be associated with particular learning algorithms.\footnote{We note that  Definition \ref{Regret_Definition} can be equivalently written as	$$\textsc{R}_{\mathcal{A}}^T=\max _{i\in A}\{\boldsymbol{\theta}_{iT}\}-\sum_{t=1}^{T}\left\langle \bold{u}_{t}, \bold{x}_{t}\right\rangle,$$
	where 	$\boldsymbol{\theta}_{iT}\triangleq \sum_{t=1}^T\bold{u}_{it}$ is the cumulative payoff for alternative $i$ until period $T$.
}

To formalize the notion that the DM implements learning algorithms that minimize the regret associated with her sequence of choices, we introduce the notion of \emph{Hannan  Consistency} (\citet{Hahn1957}).

\begin{definition}[\citet{Hahn1957}]\label{Hannan_Consistency}  A sequence $\mathcal{A}$ is Hannan Consistent if the regret in Eq. (\ref{Regret_eq}) is small, i.e., 
\begin{equation}\label{Hannan_consistency}
	\textsc{R}_{\mathcal{A}}^T=o(T).
\end{equation}
\end{definition}
Intuitively, Definition \ref{Hannan_Consistency} establishes that  the sequence $\mathcal{A}$ is Hannan consistent if the \emph{averaged} regret associated to $\mathcal{A}$ goes to zero as $T\longrightarrow \infty$. In formal terms, this  is equivalent to say  that $\mathcal{A}$ is Hannan consistent if the $\textsc{R}_{\mathcal{A}}^T$ is sublinear in $T$, i.e., if Eq. (\ref{Hannan_consistency}) holds. Thus, Hannan consistency is equivalent to saying that a sequence of choices performs as well as the best-fixed strategy in hindsight $\bold{x}^*$. Alternatively, when a sequence of choices $\mathcal{A}$ satisfies the condition (\ref{Hannan_Consistency}),  we say that the algorithm $\mathcal{A}$ satisfies the \emph{no-regret} property.\footnote{Throughout this paper, we use the terms Hannan consistency and no-regret learning interchangeably. }
\smallskip

Coupling the  ODP model with the notion of regret,  we can develop learning algorithms that satisfy Definition \ref{Hannan_Consistency}. In doing so,  most of the Game Theory and the Online Convex Optimization (OCO) literature has focused on the study of the  FTRL  algorithm using the entropic penalty (see \citet{Hazan2017} and references therein). 
\smallskip

This paper shows that a large class of discrete choice models are Hannan consistent. Our results extend the scope of no-regret learning analysis far beyond the traditional MNL model. As a consequence, no-regret learning can be studied with a richer class of behavioral models.
\subsection{The  Social Surplus Algorithm}  In this section, we develop a simple algorithm combining the ODP approach with the theory of RUMs (\citet[Ch. 5]{mcf1}). The combination of the ODP and the RUM approaches allows us to extend \citet{Hahn1957}'s approach. Formally, Hannan studies an ODP model in which the  DM's choice is given by:
\begin{equation}\label{FTPL1}
\tilde{\bold{x}}_{t+1}\in\arg\max_{\bold{x}\in\Delta_N}\langle \boldsymbol{\theta}_{t}+\eta\epsilon_{t+1},\bold{x}\rangle\quad\mbox{for $t=1,\ldots, T$},
\end{equation}
where $\boldsymbol{\theta}_t$ is the cumulative payoff vector until period $t$,    $\epsilon_{t+1}=(\epsilon_{1t+1},\ldots,\epsilon_{Nt+1})$ is a random preference shock vector, and $\eta$ is a strictly positive parameter, i.e.,  $\eta>0$. 
\smallskip

The expression (\ref{FTPL1}) establishes that the choice $\tilde{\bold{x}}_{t+1}$ is the solution of a recursive problem that depends on the past information  contained in the cumulative payoff vector $\boldsymbol{\theta}_t$, the realization of $\epsilon_{t+1}$, and  $\eta$, which is interpreted  as a learning parameter. Alternatively, the parameter $\eta$ is a measure of \emph{accuracy} on the DM's choices.\footnote{The interpretation of $\eta$ as a measure of accuracy is also considered in \citet{cerreiavioglio2021multinomial} in a  different framework.} 
\smallskip

Intuitively, the interpretation of (\ref{FTPL1}) is that the DM samples a random realization of $\epsilon_{t+1}$ to smooth out her optimization problem. In the OCO literature, this approach is called Follow the Perturbed Leader  (FTPL). A common assumption in this framework is to assume that $\epsilon_{t+1}$ is sampled from a $N$-dimensional uniform distribution. The main advantage of the FTPL idea is the possibility of inducing stability in the DM's choices.\footnote{ In the  Game Theory literature, this approach is known as fictitious stochastic play. For details, we refer the reader to \citet{FudenbergLevina1998}. } In fact,  \citet{Hahn1957} shows that when $\epsilon_t$ is i.i.d. following a  uniform distribution,  the sequence generated by  problem (\ref{FTPL1}) satisfies Definition \ref{Hannan_Consistency}. However,  the regret analysis of this approach relies on probabilistic arguments about the stochastic structure of $\epsilon_{t+1}$ rather than having a general framework (\citet{Abernethyetal2016}). More importantly,  from an economic standpoint, the interpretation of the FTPL approach is unclear, which makes it difficult to provide a behavioral foundation for the repeated stochastic choice problem.
 \smallskip

In this section, we propose that instead of focusing on particular realizations of $\epsilon_{t+1}$, we can exploit the entire distribution of it. In particular, under reasonably general distributional assumptions on $\epsilon_{t+1}$, we can use the theory of RUMs to generate an alternative approach to Hannan's original  FTPL algorithm idea. In doing so, throughout the paper, we use the following assumption.
\begin{assumption}\label{Shocks_Assumption} For all $t\geq 1$ the random vector $\epsilon_{t}=(\epsilon_{1t},\ldots,\epsilon_{Nt})$  follows a joint distribution $F=(F_{1},\ldots,F_{N})$ with zero mean that is absolutely continuous with respect to the Lebesgue measure, independent of $t$ and  $\boldsymbol{\theta}_t$,  and fully supported on $\RR^N$. 
\end{assumption}
Assumption \ref{Shocks_Assumption} is standard in random utility and discrete choice models (\citet[Ch, 5]{mcf1}). The   full support and absolute continuity  conditions imply that   (\ref{FTPL1}) can be rewritten as:
 \begin{equation}\label{FTPL2}
 \max_{\bold{x}\in\Delta_N}\langle \boldsymbol{\theta}_{t}+\eta\epsilon_{t+1},\bold{x}\rangle=\max_{j\in A}\{\boldsymbol{\theta}_{jt}+\eta\epsilon_{jt+1}\}\quad \mbox{for $t=1,\ldots,T$}.
 \end{equation}
 
 From (\ref{FTPL2}), it follows that Assumption \ref{Shocks_Assumption} implies that  $\tilde{\bold{x}}_{t+1}$  corresponds to  a corner solution, i.e., the DM chooses one of the alternatives with probability one. Furthermore, noting that $\max\{\cdot\}$ is a convex function and defining  $\tilde{\varphi}(\boldsymbol{\theta}_t+\eta\epsilon_{t+1})\triangleq \max_{j\in A}\{\boldsymbol{\theta}_{jt}+\eta\epsilon_{jt+1}\}$,  $\tilde{\bold{x}}_{t+1}$ is characterized as
\begin{equation}\label{FTPL3}
\tilde{\bold{x}}_{t+1}\in \partial \tilde{\varphi}(\boldsymbol{\theta}_{t}+\eta\epsilon_{t+1})
\end{equation}
Thus  an optimal solution $\tilde{\bold{x}}_{t+1}$ is given by a subgradient of $\tilde{\varphi}(\boldsymbol{\theta}_t+\eta\epsilon_{t+1})$ (\citet{Rockafellar1970}).  
\smallskip

Characterization (\ref{FTPL3}) is derived under the assumption of a \emph{single} realization of $\epsilon_{t+1}$.  However, a natural extension of this result is to consider the entire distribution of $\epsilon_{t+1}$ by considering the expectation of $\tilde{\varphi}(\boldsymbol{\theta}_t+\epsilon_{t+1})$. Formally,  define the function $\varphi:\RR^N\mapsto \RR$ as
\begin{equation}\label{SS_Function}
\varphi(\boldsymbol{\theta}_{t})\triangleq\EE_{\epsilon}\left(\max_{i\in A}\{ \boldsymbol{\theta}_{it}+\eta \epsilon_{it+1}\}\right).
\end{equation}

 In the RUM literature $\varphi(\boldsymbol{\theta}_t)$ is known as the \emph{social surplus} function,  which summarizes the effect of  $F$. More importantly, $\varphi(\boldsymbol{\theta}_t)$ is convex and differentiable in $\RR^N$. This latter property implies that  the choice probability vector $\bold{x}_{t+1}$  can be characterized as:
\begin{equation}\label{SS_Function_gradient}
\nabla\varphi(\boldsymbol{\theta}_t)=\bold{x}_{t+1}\quad\mbox{for $t=1,\ldots T-1$.}
\end{equation}

The previous result follows from the well-known Williams-Daly-Zachary theorem (see \citet[p. 3104]{Rust1994}).
 \smallskip
 From an economic standpoint,  the social surplus function allows us to interpret the DM's choice in terms of the theory of  RUMs. To see this,  the cumulative payoff vector $\boldsymbol{\theta}_t$ can be naturally interpreted as an \emph{estimate} of the unknown utility   $\bold{u}_t$ at time $t$. In particular, $\boldsymbol{\theta}_{jt}$ provides cumulative information about alternative $j$'s past performance. Accordingly, the random variable $\epsilon_{jt+1}$ is interpreted as a preference shock that affects how the DM  perceives the cumulative payoff associated with this particular alternative. The same logic applies to all alternatives in $A$. Thus, the DM's stochastic choice is consistent with RUMs. More importantly, this connection clarifies that different distributional assumptions on $\epsilon_{t+1}$ will imply stochastic choice rules capturing different behavioral aspects. This feature allows one to study choice models in which the alternatives exhibit high similarity and correlation. For instance, we can study models like probit and NL. Given the connection between RUMs and the ODP described above, we denote the resulting model as the RUM-ODP model.
 \smallskip
 
 A second important implication of expression (\ref{SS_Function_gradient}) is  that $\bold{x}_{t+1}$ can be interpreted as the expected value of $\tilde{\bold{x}}_{t+1}$.  The next proposition formalizes this fact.
 \begin{proposition}\label{SS_Equiv}Let Assumption \ref{Shocks_Assumption} hold. Then, for $t=1,\ldots,T$ $$\EE(\tilde{\bold{x}}_{t+1})=\bold{x}_{t+1}.$$ 
 	\end{proposition}

  From the previous result,  it follows that 
\begin{eqnarray}\label{FTPL_Potential}
\nabla\varphi(\boldsymbol{\theta}_t)&=&\EE\left(\arg\max_{\bold{x}\in\Delta_n}\langle\boldsymbol{\theta}_t+\eta\epsilon_{t+1},\bold{x}\rangle\right),\\
&=&\left(\PP\left(i=\arg\max_{j\in A}\{\boldsymbol{\theta}_{jt}/\eta+\epsilon_{t+1}\}\right)\right)_{i\in A},\nonumber\\
&=& \left(\PP(\boldsymbol{\theta}_{it}/\eta+\epsilon_{it}\geq\boldsymbol{\theta}_{jt}/\eta+\epsilon_{jt}\quad\forall j\neq i)\right)_{i\in A}.\nonumber
\end{eqnarray}

Furthermore, from the definition of  the social surplus, it is easy to see that $\varphi(\boldsymbol{\theta})=\eta\varphi(\boldsymbol{\theta}/\eta)$ and $\nabla\varphi(\boldsymbol{\theta})=\nabla\varphi(\boldsymbol{\theta}/\eta)$. We shall use this relationship in deriving our results.
\smallskip

More importantly, the convex structure of the function $\varphi$ enables us to develop a learning algorithm called the \emph{Social Surplus Algorithm} (SSA). 
In doing so, we assume that the Hessian of $\varphi(\boldsymbol{\theta}_t)$ satisfies the following technical condition.
\begin{assumption}\label{Gradient_LL} For all $t\geq 1$ the Hessian of the Social Surplus function $\nabla^2\varphi(\boldsymbol{\theta}_t)$ satisfies the following condition:
	$$2Tr(\nabla^2\varphi(\boldsymbol{\theta}_t))\leq {L\over \eta},$$
	with $L>0.$
\end{assumption}
The previous assumption is less standard as it imposes a condition on the trace of Hessian of $\varphi(\boldsymbol{\theta}_t)$. This requirement allows us to establish the Lipschitz continuity  of $\nabla\varphi(\boldsymbol{\theta}_t)$. 
\begin{lemma}\label{Gradient_Lipschitz}Let Assumptions \ref{Shocks_Assumption} and \ref{Gradient_LL} hold. Then  $\varphi(\boldsymbol{\theta})$ has a gradient-mapping that is Lipschitz continuous with constant  ${L\over \eta}$:
	$$\|\nabla \varphi(\boldsymbol{\theta})-\nabla \varphi(\tilde{\boldsymbol{\theta}})\|_1 \leq {L\over \eta}\|\boldsymbol{\theta}-\tilde{\boldsymbol{\theta}}\|_1, \quad \forall\boldsymbol{\theta} ,\tilde{\boldsymbol{\theta}}\in \RR^N,$$
	with $L>0$.
\end{lemma}
Two remarks are in order. First, as we noted before, the social surplus function $\varphi(\boldsymbol{\theta})$ can be equivalently written as $\eta\varphi(\boldsymbol{\theta}/\eta)$.  Using this equivalence, Lemma \ref{Gradient_LL} can be equivalently stated as: the social surplus function $\eta\varphi(\boldsymbol{\theta}/\eta)$ has a  gradient-mapping that is $L$-Lipschitz continuous, i.e., 

$$\|\nabla \varphi(\boldsymbol{\theta}/\eta)-\nabla \varphi(\tilde{\boldsymbol{\theta}}/
\eta)\|_1\leq  L \|\boldsymbol{\theta}/\eta-\tilde{\boldsymbol{\theta}}/\eta\|_1, \quad \forall\boldsymbol{\theta} ,\tilde{\boldsymbol{\theta}}\in \RR^N,$$
with $L>0$.
\smallskip

Our second observation is related to the fact that despite being a technical requirement, Assumption \ref{Gradient_LL} is satisfied by many well-known RUMs, like the MNL, NL, and the GEV class, as we show in  \S\ref{s33}.
\smallskip

Now  we are ready to introduce the  SSA as follows:

\begin{algorithm}
	\caption{Social Surplus Algorithm}\label{alg:euclid1}
	\begin{algorithmic}[1]
		\State Input: $\eta>0$, $F$ a distribution on $\RR^N$, and $\Delta_N$.
		\State Let $\boldsymbol{\theta}_0=0$ and choose  $\bold{x}_1=\nabla\varphi(\bold{0})$ 
		\State $\bold{for}$ $t=1$ to $T$ \textbf{do}
		\State The DM chooses $\bold{x}_{t}=\nabla\varphi(\boldsymbol{\theta}_{t-1})$
		\State The environment reveals  $\bold{u}_{t}$
		\State The DM receives the payoff $\langle\bold{u}_{t}, \nabla\varphi(\boldsymbol{\theta}_{t-1})\rangle$
		\State Update $\boldsymbol{\theta}_{t}=\bold{u}_t+\boldsymbol{
		\theta}_{t-1}$ and choose
		$$\bold{x}_{t+1}=\nabla\varphi(\boldsymbol{\theta}_t)$$
		\State \textbf{end for}
	\end{algorithmic}
\end{algorithm} 
 
The previous algorithm exploits the social surplus function and its gradient. In this sense, our algorithm is similar to the \emph{Potential Based Algorithm} proposed in \citet{BianchiLugosi2003, CesaBianchiLugosi2006}  and to the \emph{Gradient Based Algorithm} studied by \citet{Abernethyetal2016}. The idea of applying a gradient descent-like approach to study repeated stochastic choice problems dates back to \cite{Blackwell_1956}.  \cite{HART_JET_200126} and \cite{HART_GEB_2003375} apply  Blackwell's approachability theorem  to  sequential decision problems. They introduce the notion of  $\ Lambda-$ strategies, which relies on the notion of potential functions.

\smallskip

A natural question is whether \cite{BianchiLugosi2003}'s results apply to the study of the RUM-ODP model. To use their results,  we must assume that the social surplus function $\varphi$  has an additive structure. In terms of our approach, this is equivalent to assume that the random variables $\epsilon_{it}$  are $i.i.d.$ for all $t$.  For instance, the MNL model satisfies this condition. However, most of the RUMs considered in the discrete choice literature (and this paper) do not satisfy this additivity requirement. Similarly, in applying the results in  \cite{HART_JET_200126} and \cite{HART_GEB_2003375}, we need to impose restrictions  on the domain of $\nabla\varphi$. In particular, their approach  requires that  the gradient of $\varphi$ vanishes  over the \emph{approachable} set.\footnote{Formally, this is equivalent to assume that $\nabla\varphi(\bold{z})=\bold{0}$ for all $\bold{z}\in \RR_{-}^N.$}  This latter condition assumes that the random shock vector $\epsilon_t$ has bounded support, which rules out the whole class of RUMs. Thus, while related,  \cite{BianchiLugosi2003}, \cite{HART_JET_200126}, and  \cite{HART_GEB_2003375}  results do not apply to the RUM-ODP model.
\smallskip

Another important  difference between  \cite{BianchiLugosi2003}, \cite{HART_JET_200126}, and  \cite{HART_GEB_2003375}   and our approach  is related to the interpretation of the potential function. In their work, the role of a  potential function is to provide a way to measure the size of the regret associated with the DM's choice at a given time. In our framework, we can interpret the social surplus function $\varphi$ in this way. But in addition, $\varphi$ measures the expected utility received at each time point associated with the probability vector $\bold{x}_t$. In order to formalize this observation, let us define  $e_{jt}(\boldsymbol{\theta}_{t-1})\triangleq \EE(\epsilon_{jt}\mid j=\arg\max_{k}\{\boldsymbol{\theta}_{kt-1}+\epsilon_{kt}\})$ for all $j\in A. $ Then using the law of iterated expectations, it follows that $\varphi(\boldsymbol{\theta}_{t-1})$ can be expressed as a weighted average:
\begin{equation}\label{Potential_Econ}
	\varphi(\boldsymbol{\theta}_{t-1})=\sum_{i=1}^{N} \bold{x}_{it}\left(\boldsymbol{\theta}_{it-1}+\eta e_{jt}(\boldsymbol{\theta}_{t-1})\right)
\end{equation}

The expression (\ref{Potential_Econ}) makes explicit the role of the random shocks in determining the value of $\varphi(\boldsymbol{\theta}_{t-1})$.  Different distributions for the random shock $\epsilon_t$  will lead to different social surplus functional forms. This fact is relevant at least for two reasons. First, the SSA allows us to study discrete choice models with arbitrary degrees of correlation (similarity) between the different alternatives. In particular, the SSA allows us to study repeated stochastic problems in environments where the DM's choices are represented by preference trees (e.g. NL  and GNL models). Second,  by exploiting the structure of discrete choice models, the SSA can be implemented using closed-form expressions for  $\nabla\varphi(\boldsymbol{\theta}_t)$. In particular, we will show that the SSA can be implemented with the GEV class, which contains the MNL as a particular case. 
\smallskip

Without further delay, we  establish the main result of this section.

\begin{theorem}\label{SS_Surplus_algorithm_regret}Let Assumptions \ref{Shocks_Assumption} and \ref{Gradient_LL} hold. Then in the SSA we have:
\begin{equation}\label{Bound1}
\textsc{R}_{SSA}^T\leq \eta\varphi(\bold{0})+\frac{L}{2\eta}Tu^2_{max}.
\end{equation}
Furthermore, setting $\eta=\sqrt{{LTu^2_{max}\over 2\varphi(\bold{0})}}$ we get  
\begin{equation}\label{Bound2}
\textsc{R}_{SSA}^T\leq u_{max}\sqrt{2\varphi(\bold{0})LT}
\end{equation}
\end{theorem}

Some remarks are in order. First, Theorem \ref{SS_Surplus_algorithm_regret}  establishes that by implementing the SSA, a large class of RUMs are Hannan consistent. The result highlights the role of  $L$, $\eta$, and the  social surplus function evaluated at $\bold{u}=\bold{0}$.\footnote{Note that in this case $\eta\varphi(\bold{0})=\eta\EE(\max_{i=1,\ldots,N}\{\epsilon_i\})$.} This result enables one to implement the SSA exploiting different functional forms for the stochastic choice rule given by $\nabla\varphi(\boldsymbol{\theta}_t)$. In  \S \ref{s33} we will show how the result  in   Theorem \ref{SS_Surplus_algorithm_regret}  applies to  a large class of RUMs.
\smallskip

Second, from a technical point of view, the proof of Theorem \ref{SS_Surplus_algorithm_regret} relies on the convex structure of $\varphi(\boldsymbol{\theta}_t).$ Formally, we exploit convex duality arguments to bound the regret of the SSA. In \S\ref{s4} we shall further exploit these results to analyze the FTRL algorithm.
\smallskip

Third,  Theorem \ref{SS_Surplus_algorithm_regret} is related to the results in \citet{BianchiLugosi2003}. Their analysis focuses on several potential functions that generate the choice probability vector $\bold{x}_{t+1}$. Our analysis differs from theirs in that $\varphi(\boldsymbol{\theta}_t)$ has an economic meaning in terms of RUMs. Similarly, it is worth pointing out that Theorem \ref{SS_Surplus_algorithm_regret} generalizes the analysis in \citet{Abernethyetal2014, Abernethyetal2016}. They focus mainly on the  MNL  model, which corresponds to the case of  $\epsilon_t$ following an extreme value type 1 distribution. The following result formalizes this observation. 

\begin{corollary}\label{Cor_Logit} Assume that  for all $t$ the random shock vector  $\epsilon_t$ follows a Extreme Value type 1 distribution, given by 
	 \begin{equation}\label{EV1}
	F(\epsilon_{1t},\ldots,\epsilon_{Nt})=\exp\left(-\sum_{j=1}^Ne^{-\epsilon_{jt}}\right)\quad\forall t.
	\end{equation}

	 Then, in the SSA, setting $\eta=\sqrt{{Tu^2_{max}\over 2 \log N}}$ yields:
$$\textsc{Regret}_T\leq u_{max}\sqrt{2\log N T}.$$
	\end{corollary}

The previous  corollary follows from  the fact that for the MNL  it is well known that (cf. \citet[Ch. 5]{mcf1} and \citet[Ch. 3]{train_2009}) $$\varphi(\boldsymbol{\theta}_t)=\eta\log\left( \sum_{j=1}^Ne^{\boldsymbol{\theta}_{jt}/\eta}\right).$$
Noting that 
 $\eta\varphi(\bold{0})=\eta\log\sum_{i=1}^Ne^{0}=\eta\log N$ and using the fact that $\nabla\varphi(\boldsymbol{\theta})=\left( {e^{\boldsymbol{\theta}_i/\eta} \over \sum_{j=1}^N e^{\boldsymbol{\theta}_j}/\eta}\right)_{i\in N}$ is ${1\over\eta}$-Lipschitz continuous, it follows that Corollary \ref{Cor_Logit} is a direct application of Theorem \ref{SS_Surplus_algorithm_regret}.
 \smallskip
 
From a behavioral standpoint, the MNL imposes that the stochastic shocks $\{\epsilon_{jt+1}\}_{j\in A}$ are independent across alternatives. This assumption can be strong, especially in environments where the payoffs associated with the options in $A$ can be correlated. The following section discusses how to relax the independence assumption to implement the SSA with more flexible RUMs.
 
 \section{The SSA and GEV models}\label{s33}
 This section aims to connect the  RUM-ODP model and the SSA approach with the class of GEV models. To the best of our knowledge, the GEV class has not been studied in the context of no-regret learning models.
  \smallskip
  
The GEV class was introduced by \citet{McFadden1978} to generalize the MNL model by allowing general patterns of dependence among the unobserved components of the options while yielding analytical and tractable analytical closed forms for the choice probabilities under consideration. 
  \smallskip
  
 In developing the GEV class, McFadden introduces the notion of a \emph{generator function}, which we define now.
 \begin{definition}\label{Generator_GEV} A  function $G:\RR^N_{+}\longrightarrow\RR_+$ is a generator if the following conditions hold:
 	\begin{itemize}
 		\item[i)] For all $\bold{y}=(\bold{y}_1,\ldots,\bold{y}_N)\in \RR^N_+$, $G(\bold{y})\geq 0$.
 		\item[ii)] The function $G$ is homogeneous of degree 1: $G(\lambda\bold{y})=\lambda G(\bold{y})$ for all $\bold{y}\in \RR^N_+$ and $\lambda>0.$
 		\item[iii)]For $i=1,\ldots,N$, $G(\bold{y})\longrightarrow\infty$  as $\bold{y}_i\longrightarrow \infty$.
 		\item[iv)]  If $i_1,\ldots,i_k$ are distinct  from each other  $k$ distincts components, then 
 		$i_1,\ldots,i_k$, the k-th order partial derivative ${\partial G(\bold{y}_{1},\ldots,\bold{y}_{N})\over \partial \bold{y}_{{i}_1}\cdots\partial\bold{y}_{{i}_k}}\geq 0$  when $k$ is odd, whereas ${\partial G(\bold{y}_{1},\ldots,\bold{y}_{n})\over \partial \bold{y}_{{i}_1}\cdots\partial\bold{y}_{{i}_k}}\leq 0$  when $k$ is even.
 	\end{itemize} 
 \end{definition}

 \citet{McFadden1978,mcf1} show that when the generator function $G$ satisfies  i)-iv), then the random vector $\epsilon=(\epsilon_1,\ldots,\epsilon_N)$ follows a Multivariate Extreme Value distribution:
  \begin{equation}\label{MEV}
 F(\epsilon_1,\ldots,\epsilon_N)=\exp\left(-G(e^{-\epsilon_1},\ldots,e^{-\epsilon_N})\right).
  \end{equation}

More importantly, McFadden shows that when $\epsilon$ follows distribution (\ref{MEV}), a random utility maximization model is consistent with the  RUM hypothesis.  Furthermore, \citet{McFadden1978} establishes  that for a deterministic utility vector $\boldsymbol{\theta}$ with $
\bold{y}=(e^{\boldsymbol{\theta}_1},\ldots, e^{\boldsymbol{\theta}_N})$, the social surplus function can be expressed in ``closed'' form expression as: 
\begin{equation}\label{GEV_Suprlus}
	\varphi(\boldsymbol{\theta})=\log G(\bold{y})+\gamma,
\end{equation}
where   $\gamma=0.57721$ is the Euler's constant.
\smallskip

Using the fact $\nabla\varphi(\bold{u})$ yields the choice probabilities, we get
\begin{eqnarray}\label{GEV_choice}
	\nabla_i\varphi(\bold{u})&=&\frac{\bold{y}_iG_i(\bold{y})}{\sum_{j=1}^N\bold{y}_jG_j(\bold{y})}\quad\mbox{for $i=1,\ldots,N$},\\
	&=&{e^{\boldsymbol{\theta}_i+\log G_i(e^{\boldsymbol{\theta}})} \over \sum_{j=1}^N e^{\boldsymbol{\theta}_j+\log G_j(e^{\boldsymbol{\theta}})}}.\nonumber
\end{eqnarray}

Eqs.   (\ref{GEV_Suprlus})  and (\ref{GEV_choice})  highlight two fundamental properties
of the GEV class. First, the choice probability vector has a logit-like form, which makes the analysis tractable.  Second, the closed form expression for $\varphi$ helps us to study the SSA in a large class of RUMs. 

\subsection{The RUM-ODP  model  and the GEV class} Now we connect the GEV class with our RUM-ODP approach. In doing so, we note that given the cumulative payoff vector $\boldsymbol{\theta}_t$ and the learning parameter $\eta$, the social surplus function can be written as:
$$\varphi(\boldsymbol{\theta}_t)=\EE\left(\max_{j\in A}\{\boldsymbol{\theta}_{jt}+\eta\epsilon_{jt+1}\}\right) =\eta\varphi(\boldsymbol{\theta}_t/\eta).$$

Then, Eqs. (\ref{GEV_Suprlus}) and (\ref{GEV_choice}) can be written as:

 \begin{equation}\label{GEV_Suprlus_Learning}
\varphi(\boldsymbol{\theta}_t)=\eta(\log G(e^{\boldsymbol{\theta}_t/\eta})+\gamma)
 \end{equation}
 and 
 \begin{eqnarray}\label{GEV_choice_Learning}
 \nabla_i\varphi(\boldsymbol{\theta}_t)&=& {e^{\boldsymbol{\theta}_{it}/\eta+\log G_i(e^{\boldsymbol{\theta}_t/\eta})} \over \sum_{j=1}^N e^{\boldsymbol{\theta}_{jt}/\eta+\log G_j(e^{\boldsymbol{
 		\theta}_{t}/\eta})}}\quad \mbox{for $i=1,\ldots,N$}.
 \end{eqnarray}
 
 It is worth remarking that expression (\ref{GEV_choice_Learning}) is derived using  the fact  that for all $i$, ${\partial \varphi(\boldsymbol{\theta }_t)\over \partial \boldsymbol{\theta}_{it}}=\eta{\partial \varphi(\boldsymbol{\theta }_t/\eta)\over \partial \boldsymbol{\theta}_{it}}.$ In deriving this relationship, we define the GEV model in terms of the \emph{scaled} utility vector $\boldsymbol{\theta}_t/\eta=\left(\boldsymbol{\theta}_{it}/\eta\right)_{i\in A}$.

\begin{example}\label{Logit_Example}Consider the linear aggregator $G(e^{\boldsymbol{\theta}_t/\eta})=\sum_{j=1}^Ne^{\boldsymbol{\theta}_{jt}/\eta}$. Then it follows that
	\begin{eqnarray}
	\varphi(\boldsymbol{
		\theta}_t)&=&\eta\varphi(\boldsymbol{\theta}_t/\eta)\nonumber\\
	&=&\eta\log G(e^{\boldsymbol{\theta}_t/\eta})+\eta\gamma\nonumber\\
	&=&\eta\log\sum_{j=1}^Ne^{\boldsymbol{\theta}_{jt}/\eta}+\eta\gamma\nonumber
	\end{eqnarray}

and 
$$\nabla_i\varphi(\boldsymbol{\theta}_t)={e^{\boldsymbol{\theta}_{it}/\eta}\over\sum_{j=1}^N e^{\boldsymbol{\theta}_{jt}/\eta}}\quad \mbox{for $i=1,\ldots,N$}.$$
\end{example}

By exploiting the properties of the GEV class, we can establish that the SSA  achieves Hannan consistency. In doing so, we use the following technical result from \citet[Thm. 3]{Nesterovetal2019}.

\begin{lemma}\label{Result_GEV} Let $\bold{y}_t=(e^{\boldsymbol{\theta}_{1t}/\eta},\ldots,e^{\boldsymbol{\theta}_{Nt}/\eta})\in \RR_+^N$ and let $G$ be  a generator function satisfying the following inequality
	\begin{equation}\label{GEV_condition}
	\sum_{i=1}^{N} \frac{\partial^{2} G(\bold{y}_t)}{\partial \bold{y}_{it}^2} \cdot (\bold{y}_{it})^{ 2} \leq M G(\bold{y}_{1t},\ldots,\bold{y}_{Nt}), \quad\mbox{for all $t=1,\ldots, T$,}
	\end{equation}
for some $M\in \RR_{++}$.
	Then the  social surplus function   $\varphi(\boldsymbol{\theta}_t)=\eta\left(\log G(\bold{y}_t)+ \gamma\right)$ has a  Lipschitz continuous  gradient with constant $L={2M+1\over \eta}.$
		
\end{lemma}

The previous lemma establishes that the Lipschitz constant $L$ depends on the parameters $M$ and  $\eta$. With this result in place, we can show the following bound on the regret for the  GEV class:
\begin{theorem}\label{SS_algorithm_GEV} Let Assumption \ref{Shocks_Assumption} hold. In addition, assume that there exists a generator function $G$ satisying Eq. (\ref{GEV_condition}). Then, in the  SSA 
	\begin{equation}\label{Regret_GEV_SSA}
		\textsc{R}^T_{SSA}\leq \eta\log G(\bold{1})+\frac{L}{2\eta}Tu^2_{max},
	\end{equation}
where $L=2M+1$.
Furthermore, setting $\eta=\sqrt{LTu^2_{max}\over 2\log G(\bold{1})}$, we get:

	\begin{equation}\label{Regret_GEV_SS}
\textsc{R}^T_{SSA}\leq u_{max}\sqrt{2 \log G(\bold{1})(2M+1)T}.
\end{equation}
\end{theorem}

The previous result provides a bound to the regret associated with the SSA when the GEV is considered. In other words, when the GEV class is combined with the RUM-ODP model, the sequence of choices generated by the SSA achieves is Hannan consistent. To the best of our knowledge, Theorem \ref{SS_algorithm_GEV} is the first regret analysis using the GEV class. In the next sections we discuss several widely used GEV models that satisfy condition (\ref{GEV_condition}).
\subsection{The Generalized Nested Logit (GNL) model} The most widely used model in the GEV class is the Generalized Nested Logit  (GNL) model (\citet{McFadden1978} and \citet{WenKoppelman2001}). This approach generalizes several GEV models like NL and Ordered GEV  (OGEV). Given its flexibility, the GNL has been applied to energy, transportation, housing, telecommunications, demand estimation, etc.\footnote{ For a discussion about several applications of GEV models, we refer the reader to \citet{train_2009} and the references therein.}  The main advantage of using a GNL is the possibility of incorporating correlation between the elements of the random shock vector  $\epsilon_{t+1}$ in a relatively simple and tractable manner.  
 \smallskip
 
 Let $A$ be the set of options  partitioned into $K$   nests labeled $\mathcal{N}_1,\ldots, \mathcal{N}_K$. Let $\mathcal{N}$ be the set of all nests. In defining the set $\mathcal{N}$, we allow for overlapping between the nests. In particular,  an option $i$ may be an element of more than one nest. For instance, an option $i$ may be an element of  nests $\mathcal{N}_k$, $\mathcal{N}_{k^\prime}$, $\mathcal{N}_{k^{\prime\prime}}$ simultaneously. 
  \smallskip
 
 For $k=1,\ldots,K$, let $0< \lambda_{k}\leq 1$ be nest-specific parameters. From an economic standpoint,  the parameter $\lambda_k$ is a  measure of the degree of independence between each random shock $\epsilon_{i}$ in nest $k$. In particular, the statistic $1-\lambda_k$ is a measure of correlation (\citet[Ch. 5]{train_2009}). Thus, as the value of  $\lambda_k$ increases,  the value of this statistic decreases, indicating less correlation. 
 \smallskip
  Given the cumulative payoff vector $\boldsymbol{\theta}_t$, the  generator function $G$  for the GNL is the following:
\begin{equation}\label{GNL_generating_function}
G(e^{\boldsymbol{\theta}_t/\eta})=\sum_{k=1}^K\left(\sum_{i=1}^{N}\left(\alpha_{i k} \cdot e^{\boldsymbol{\theta}_{it}/\eta}\right)^{1 / \lambda_{k}}\right)^{\lambda_{k}}.
\end{equation}

The parameter  $\alpha_{ik} \geq 0$  characterizes the ``portion'' of the alternative $i$ assigned to nest $k$. Thus for each   $i \in A$ the allocation parameters must satisfy:
$$\sum_{k=1}^K \alpha_{ik}=1.$$

Using the previous condition, the set of alternatives within the $k$-th nest is defined as:
$$\mathcal{N}_{k}=\left\{i\in A \mid \alpha_{ik}>0\right\},$$
where $A=\bigcup_{k=1}^K\mathcal{N}_k$.

Based on the previous description, it is easy to show that the  function $G$ in Eq. (\ref{GNL_generating_function}) defines a GEV model (cf. \citet{WenKoppelman2001}). 
\smallskip

 To describe the choice probability vector $\bold{x}_{t+1}$ to be used in the SSA, we note that the GNL is a model in which the underlying choice process comprises two stages. In the first stage, the DM chooses  nest $k$ with  probability 
\[
\PP_{kt}=\frac{e^{\bold{v}_{kt} }}{\sum_{\ell=1}^K e^{\bold{v}_{\ell t} }}
\]
where
\[
\bold{v}_{kt}=\lambda_{k} \log \left(\sum_{i=1}^{N}\left(\alpha_{ik} \cdot e^{\boldsymbol{\theta}_{it}/\eta}\right)^{1 / \lambda_{k}}\right)
\]
stands for the \emph{inclusive} value  contained in  nest $k$.
\smallskip

In the second stage, the probability of choosing  alternative $i$ within nest $k$ is given by:
\[
\PP_{ikt}=\frac{\left(\alpha_{ik} \cdot e^{\boldsymbol{\theta}_{it}/\eta}\right)^{1 / \lambda_{k}}}{\sum_{j=1}^{N} \left(\alpha_{jk} \cdot e^{\boldsymbol{\theta}_{jt}/
	\eta}\right)^{1 / \lambda_{k}}}.
\]

Thus, according to the GNL,  the probability of  choosing alternative $i$ is expressed as $\nabla_i\varphi(\boldsymbol{\theta}_t)=\bold{x}_{it+1}$, where
\[
\bold{x}_{it+1}=\sum_{k=1}^K \PP_{kt} \cdot \PP_{ikt}\quad\mbox{for $i=1,\ldots,n,$ $t=1,\ldots,T-1.$}
\]

Lemma \ref{GNL_regret_lemma} in Appendix \ref{Proofs}  establishes two important properties of the  GNL model. First, it shows that  $\nabla\varphi(\boldsymbol{\theta}_t)$ is Lipschitz  continuous with constant ${{2\over \min_{k}\lambda_k}-1\over \eta}$.   The second important property is that  $\log G(\bold{1})\leq \log N.$  These two observations allows us to establish the following result.

\begin{proposition}\label{GNL_regret} Let  $0<\lambda_k\leq 1$ for $k=1,\ldots,K$.  In addition, set  $\eta=\sqrt{\left( {2\over \min_{k}\lambda_k}-1\right)Tu^2_{max}\over 2\log N}$ . Then in  the GNL we have
	\begin{equation}\label{Bound2_GEV}
	\textsc{R}_{SSA}^T\leq u_{max}\sqrt{2\log N \left( {2\over \min_{k}\lambda_k}-1\right)T}.
	\end{equation}

\end{proposition}

Some remarks are in order. First, Proposition \ref{GNL_regret} is a direct corollary of  Theorem \ref{SS_algorithm_GEV}. It establishes that in the RUM-ODP model, the sequence of choices generated by the GNL model is  Hannan consistent. 
\smallskip

Second, Proposition \ref{GNL_regret} expands the scope of the SSA  to more general environments than those considered by the traditional MNL model. The fundamental object in this result is the specification of the generator function  $G$. In Appendix \ref{AppendixGNL}, we discuss how different specifications of the GNL model yield the  Paired Combinatorial logit (PCL), the  OGEV model, and the Principles of Differentiation GEV model.  
\smallskip

Finally, it is worth pointing out that   Proposition \ref{GNL_regret} is related to the Potential-Based Algorithm introduced by  \citet{BianchiLugosi2003} and further explored by \citet{Abernethyetal2014,Abernethyetal2016}. In particular,   the generator $G(e^{\boldsymbol{\theta}_t/\eta})$  can be seen as a potential function  that generates the choice prediction  vector $\bold{x}_{t+1}$. Thus Proposition \ref{GNL_regret}  can be seen as a way to extend  \citet{BianchiLugosi2003} and \citet{Abernethyetal2014,Abernethyetal2016} approach to RUMs far beyond the MNL model.
\subsection{The Nested Logit (NL) model}\label{Applications_GNL} 
The NL model proposed by \citet{McFadden1978} as a particular case of the GNL in the sense that each alternative $i\in A$ belongs to a unique nest.\footnote{In addition, in Appendix \ref{AppendixGNL}  we discuss the CNL, OGEV, and the PDGEV models.} In other words, the NL is a model in which the nests are 
\emph{mutually exclusive}. In particular,  each allocation parameter  $\alpha_{ik}$ is: 
$$\alpha_{ik}=\begin{cases} 1 &\mbox{if alternative $i\in \mathcal{N}_k$ }  \\ 
	0& \mbox{otherwise}  \end{cases} .
$$

Accordingly,  the  generator $G$ corresponds to:
$$G(e^{\boldsymbol{\theta}_t/\eta})=\sum_{k=1}^K\left(\sum_{i\in \mathcal{N}_k}(e^{\boldsymbol{\theta}_{it}/\eta})^{	1/\lambda_k}\right)^{\lambda_k}.$$

In this model the gradient $\nabla\varphi(\boldsymbol{\theta})$ is  Lipschitz continuous with constant $
\left({2\over\min\lambda_k}-1\right) /\eta$. Thus, by Proposition \ref{GNL_regret} we conclude that the NL model is Hannan consistent.
\smallskip

It is worth mentioning that when $\lambda_k=1$ for $k=1,\ldots,K$, the NL boils down to the MNL.\footnote{We recall that when $\lambda_k=1$ for $k=1,\ldots,N$  the elements of the  random vector $\epsilon_t$ are independent. } Under this parametrization, the generator $G$ can be written as
\begin{eqnarray}
	G(e^{\boldsymbol{\theta}_t/\eta})&=&\sum_{k=1}^K\sum_{i\in \mathcal{N}_k}e^{\boldsymbol{\theta}_{it}/\eta},\nonumber\\
	&=&\sum_{i=1}^Ne^{\boldsymbol{\theta}_{it}/\eta}.\nonumber
\end{eqnarray}
Thus, as a direct result of Proposition \ref{GNL_regret}, we find that the MNL is Hannan consistent.

\section{ The RUM-ODP model and the FTRL algorithm}\label{s4}
As described in the introduction section, the FTRL algorithm models the  DM  as solving a strictly concave optimization problem combining past information (cumulative payoffs) and a  deterministic regularization term. Intuitively, a  regularization function can be seen as a  hedging mechanism against bad future events or as a way of avoiding overfitting the observed data (\citet{Shalev-Shwartz2012}). Thus, at each point in time, the solution to this regularized optimization problem yields the  DM's optimal probability distribution as a function of cumulative payoffs.
\smallskip

In this section, we provide an economic foundation for the FTRL algorithm. Formally,  we establish the connection between the  RUM-ODP, the SSA, and the  FTRL algorithm. We make three contributions. First,  we exploit the convex structure of the social surplus function to derive a strongly convex regularization function denoted by $\mathcal{R}(\bold{x})$. Formally, $\mathcal{R}(\bold{x})$ corresponds to the convex conjugate of $\varphi(\boldsymbol{\theta})$.  As a direct consequence of this convex conjugacy, we show that the SSA and the FTRL algorithms are equivalent from a convex analysis point of view. In addition, we show that the choice probability vector generated by the FTRL algorithm can be expressed in a general recursive form. This latter fact implies that our approach generalizes in a nontrivial way the EWA, which only applies to the MNL model. Second, we show that in the case of the NL model, $\R(\bold{x})$ can be expressed in closed form. Finally, we show how by using the FTRL approach, the RUM-ODP  model can incorporate the  \emph{recency bias} effect.  
\subsection{Convex conjugate and regularization} From \S\ref{s33} we know that $\varphi(\boldsymbol{\theta})$ is convex and differentiable. Following \citet{Rockafellar1970}, we define the convex conjugate of  $\varphi(\boldsymbol{\theta})$, denoted by $\varphi^*(\bold{x})$ as:
 \begin{equation}
 \varphi^*(\bold{x})=\sup_{\boldsymbol{\theta}\in 
	\RR^N}\{\langle \boldsymbol{\theta},\bold{x}\rangle-\varphi(\boldsymbol{\theta})\}.
 \end{equation}
Using the fact  that $\varphi(\boldsymbol{\theta})=\eta\varphi(\boldsymbol{\theta}/\eta)$  combined with \citet[Thm. 4.14(b)]{Beck2017}  we can define  the function $\mathcal{R}(\bold{x})=\eta\varphi^*(\bold{x};\eta)$, where $\varphi^*(\bold{x};\eta)$ is the convex conjugate of the \emph{parametrized} social surplus function $\varphi(\boldsymbol{\theta}/\eta)$.  The next result summarizes some fundamental properties of $\mathcal{R}(\bold{x})$. 
\begin{proposition}\label{R_smooth}Let Assumptions \ref{Shocks_Assumption} and \ref{Gradient_LL} hold. Then:
	\begin{itemize}
		\item[i)] $\mathcal{R}(\bold{x})$  is  ${\eta\over L}$-strongly convex.
		\item[ii)] $\mathcal{R}(\bold{x})$  is differentiable  for all $\bold{x}\in int \Delta_N$. 
		\item[iii)] The optimization problem
		\begin{equation}\label{Opt_Program}
			\max_{\bold{x}\in \Delta_N}\{\langle\boldsymbol{\theta},\bold{x}\rangle-\mathcal{R}(\bold{x})\}
		\end{equation}
		has a unique solution. Furthermore:
		$$\nabla\varphi(\boldsymbol{\theta})=\arg\max_{\bold{x}\in \Delta_N}\{\langle\boldsymbol{\theta},\bold{x}\rangle-\mathcal{R}(\bold{x})\}.$$
		\end{itemize}
\end{proposition}
Some remarks are in order. First, part i) follows from a fundamental equivalence between  the Lipschitz continuity of $\varphi(\boldsymbol{\theta})$  and the strong convexity of its convex conjugate $\varphi^*(\bold{x})$. This equivalence is known as the Baillon-Haddad Theorem; see, e.g., \citet[12, Section H]{RockafellarWets1997}  and \citet{BauschkeCombettes2010}.\footnote{ Lemma \ref{Baillon_Haddad} in Appendix \ref{Proofs} provides the formal statement of this result.} Second, Proposition \ref{R_smooth}ii) establishes that $\R(\bold{x})$ is differentiable so that we can exploit the first order conditions to find the optimal $\bold{x}^*$. Third, Proposition \ref{R_smooth}iii) follows from the Fenchel equality. The main implication of this part is that given the cumulative payoff $\boldsymbol{\theta}_t$, the choice probability vector $\bold{x}_{t+1}$ can be equivalently characterized as the unique solution of a strongly concave optimization program. This latter fact allows us to provide a simple interpretation to  $\R(\bold{x})$. In particular, for $\bold{x}_{t}=\nabla\varphi(\boldsymbol{\theta}_{t-1})$, and using Eq. (\ref{Potential_Econ}) combined with the Fenchel equality, we find that 
	\begin{equation}\label{Reg_Weighted}
		\R(\bold{x}_{t})=-\eta\sum_{j=1}^N\bold{x}_{jt}e_{jt}(\boldsymbol{\theta}_{t-1})\quad\mbox{for $t\geq 1$.}
\end{equation}

The expression (\ref{Reg_Weighted}) makes explicit the role of the distribution of $\epsilon_t$  in determining the shape of the regularizer.  For instance, for the the MNL model, it is well known that $e_{jt}(\boldsymbol{\theta}_{t-1})=-\log \bold{x}_{jt}$ for all $j\in A$. In this case we obtain $\R(\bold{x}_t)=\eta\sum_{j=1}^N\bold{x}_{jt}\log\bold{x}_{jt}$, which is just the familiar entropic regularization. As we shall see in \S\ref{NL_Logit_Form},  in applying similar arguments to the case of the NL model, we can also provide a closed form expression for $\R(\bold{x}_t).$
\smallskip

From a technical point of view, \citet{Sandholm2002}, \citet{Abernethyetal2014, Abernethyetal2016}, \citet{GuiyunLiZizhuo2017}, and \citet{Fosgerauetal2019a} establish a similar equivalence as the one in (\ref{Opt_Program}). The proof of our result borrows some of their arguments. We contribute to  their results by adding the property of strong convexity of  $\R(\bold{x})$. 
\subsection{The FTRL algorithm}
The main implication  of Proposition \ref{R_smooth} is that $\R(\bold{x})$ can be seen as a regularization function (\citet{Hazan2017}). This result implies that introducing  a stochastic pertubation $\epsilon_t$ is equivalent to introducing a deterministic regularization $\mathcal{R}(\bold{x})$. More importantly,  Proposition \ref{R_smooth} allows one to write the   FTRL algorithm as follows:

\begin{algorithm}
	\caption{Follow the regularized  leader}\label{alg:euclid2}
	\begin{algorithmic}[1]
		\State Input: $\eta>0$, $\mathcal{R}$, and $\Delta_N$.
		\State Let $\bold{x}_1=\arg\max_{\bold{x}\in\Delta_N}\left\{-\mathcal{R}(\bold{x})\right\}$.
		\State $\bold{for}$ $t=1$ to $T$ \textbf{do}
		\State Predict $\bold{x}_t$.
		\State The environment reveals $\bold{u}_t$.
		\State The DM receives the payoff $\langle\bold{u}_t,\bold{x}_t\rangle$. 
		\State Update $\boldsymbol{\theta}_t=\bold{u}_t+\boldsymbol{\theta}_{t-1}$ and choose
		$$\bold{x}_{t+1}=\arg\max_{\bold{x}\in \Delta_N}\left\{\langle\boldsymbol{\theta}_t,\bold{x}\rangle-\mathcal{R}(\bold{x})\right
		\}.$$
		\State \textbf{end for}
	\end{algorithmic}
\end{algorithm}

Using the FTRL approach, we can establish the Hannan consistency of a large class of discrete choice models. 
 
\begin{theorem}\label{No_regret_Discrete_choice}Let Assumptions \ref{Shocks_Assumption} and \ref{Gradient_LL} hold. Then the  FTRL Algorithm satisfies the following bound
	$$	\textsc{R}_{FTRL}^T\leq \eta\varphi(\bold{0})+{L\over 2\eta} Tu^2_{\text{max}}.$$
Furthermore, setting $\eta=\sqrt{{LTu^2_{max}\over 2\varphi(\bold{0})}}$ we get  
\begin{equation*}
	\textsc{R}_{FTRL}^T\leq u_{max}\sqrt{2\varphi(\bold{0})LT}.
\end{equation*}	
\end{theorem}

Three remarks are in order. First, Theorem \ref{No_regret_Discrete_choice}  establishes that the FTRL algorithm achieves the Hannan consistency property. This result is similar to the conclusion we obtained in Theorem \ref{SS_Surplus_algorithm_regret}. Its proof exploits the convex duality structure of the social surplus function. Thus, the  SSA and the FTRL algorithm are \emph{dual} to each other. This result implies any of the two algorithms achieves Hannan consistency. 
\smallskip

Second, we point out that in obtaining the result in Theorem \ref{No_regret_Discrete_choice}, no knowledge about the functional form of $\mathcal{R}(\bold{x})$ is required. This feature is not new in the analysis of the FTRL algorithm. However, thanks to the RUM-ODP model's convex structure, we can use the information contained in the choice probability vector to perform our regret analysis. Thus, from a behavioral standpoint, the FTRL approach can be interpreted as an algorithm that allows learning in the case of perturbed random utility models (cf. \citet{Sandholm2002}, \citet{Fudenbergetal2015}, and \citet{Fosgerauetal2019a}). More importantly, our result shows that the FTRL algorithm has an economically grounded optimization interpretation.
\smallskip

We close this section by formalizing the equivalence between the FTRL and SSA approaches. 
\begin{theorem}\label{Equiv_SSA_FTRL} Let Assumptions \ref{Shocks_Assumption} and \ref{Gradient_LL} hold. Then the SSA and the FRTL algorithms are equivalent.
\end{theorem}
The proof of the previous theorem relies on  Proposition \ref{R_smooth}. In simple, Theorem \ref{Equiv_SSA_FTRL} establishes that the  FTRL algorithm and the SSA approach are \emph{dual} to each other. From a technical standpoint, the assumption that $\varphi(\boldsymbol{\theta}_t)$ has a Lipschitz continuous gradient is key in deriving this equivalence. As discussed in \S \ref{s33}, a large class of GEV satisfies this condition. Thus, the FTRL algorithm is useful for studying no-regret learning algorithms in cases far beyond the MNL case.
\subsection{A general recursive structure}\label{Recursive_Structure}
In the previous sections  we have defined $\mathcal{R}(\bold{x})$  as  the convex conjugate of $\varphi(\boldsymbol{\theta})$. In particular, we discussed how specific assumptions on the distribution of $\epsilon_t$ lead to different functional forms for $\mathcal{R}(\bold{x})$. This section aims to show how under Assumptions \ref{Shocks_Assumption} and \ref{Gradient_LL}, a general recursive structure can is available for the choice probability vector $\bold{x}_t$.  The main appeal of this recursive structure is that it does not require knowledge of the functional form associated with $\R(\bold{x})$.
\smallskip

We begin noticing that the choice probabilities are given by:
	\begin{equation}\label{Grad_logit}
		\bold{x}_{it+1}={H_{i}(e^{\boldsymbol{\theta}_t/\eta})\over \sum_{j=1}^NH_{j}(e^{ \boldsymbol{\theta}_{t}/\eta})}\quad\mbox{for all $i\in A$, $t\geq 1$},
\end{equation}where the vector-valued function \(H(\cdot)=\left(H_{j}(\cdot)\right)_{j=1,\ldots,N}: \mathbb{R}_{+}^{n} \mapsto \mathbb{R}_{+}^{n}\) is defined
	as the gradient of the exponentiated surplus, i.e.
	\begin{equation}\label{Grad_Exp}
		H\left(e^{\boldsymbol{\theta}_{t}/\eta}\right)=\nabla\left(e^{\varphi\left(\boldsymbol{\theta}_{t}\right) }\right).
	\end{equation}

Two remarks are in order. First, the derivation in  Eqs.(\ref{Grad_logit})-(\ref{Grad_Exp}) were proposed  by \citet{Fosgerauetal2019a} as a way to characterize the  probability vector in the case of discrete choice models satisfying Assumptions \ref{Shocks_Assumption}. In particular,  Eq.(\ref{Grad_logit}) is a straightforward application of their results, showing that the choice probabilities have a logit-like form. Second, from \citet[Prop. 2]{Fosgerauetal2019a} it follows  that  the vector-valued function $H(\cdot)$ is globally invertible. Exploiting this property,   we  can define  $\Phi(\cdot)\triangleq H^{-1}(\cdot)$. Combining the previous definition with  \citet[Prop. 3ii)]{Fosgerauetal2019a},   it follows  that  $\R(\bold{x}_t)=\eta\langle\bold{x}_t,\log \Phi(\bold{x}_t)\rangle$.  Based on these observations, we  establish the following result:

\begin{proposition}\label{Recursive_Choice_FTRL} Let Assumption \ref{Shocks_Assumption} and \ref{Gradient_LL} hold. Then in  the FTRL algorithm  we have:
	\begin{equation}\label{Recursive_Prob}
		\bold{x}_{it+1}={H_i(e^{\bold{u}_t/\eta+\alpha(\bold{x}_t)})\over \sum_{j=1}^NH_j(e^{\bold{u}_t/\eta+\alpha(\bold{x}_t)})} \quad \forall i\in A, t\geq 1
	\end{equation}
	with $\alpha(\bold{x}_t)\triangleq \log \Phi(\bold{x}_t).$
\end{proposition}
Some remarks are in order. First,  Proposition \ref{Recursive_Choice_FTRL} provides a recursive expression for the choice probabilities at each period.  Intuitively, Eq. (\ref{Recursive_Prob}) shows that in the FTRL  algorithm, the DM   incorporates
the past information to choose the vector $\bold{x}_{t+1}$ through the term $\alpha(\bold{x}_t)$. In particular,  the term $\alpha(\bold{x}_t)$ is a weight  that increases  (decreases) the choice probability of those alternatives that have had better (worse) payoffs in the past. To see this,  we note that for each alternative $i\in A$, the associated payoff can be written as $\bold{u}_{it}+\alpha_i(\bold{x}_t)$ for $t\geq 1$. Thus, the term $\alpha(\bold{x}_{t})$ contains the past information about the performance of the different alternatives. A  second observation is related to the fact that Proposition \ref{Recursive_Choice_FTRL} makes explicit that different assumptions on $\epsilon_t$ will generate different choice probability vectors. In particular, the functional form of $\Phi(\bold{x}_t)$ is determined by the  distribution of $\epsilon_t.$ The argument behind this fact comes from  \cite[Prop. 2]{Fosgerauetal2019a}, which implies that   $-\log\Phi_j(x_t)=\varphi(\boldsymbol{\theta}_t)-\boldsymbol{\theta}_{jt}$ for all $j\in A, t\geq 1.$
\subsection{NL and a closed form expression for $\mathcal{R}(\bold{x})$}\label{NL_Logit_Form} Formally, when implementing the FTRL algorithm, a fundamental question is how to choose the regularization $\mathcal{R}(\bold{x})$. In the previous section  we defined  $\mathcal{R}(\bold{x})$  as  the convex conjugate of $\varphi(\boldsymbol{\theta})$.  However, there is no available closed form for this regularization term for general distributions of the random vector $\epsilon_t$. This feature is common in the theory of online learning problems, where the Euclidean and entropic penalty terms are the two commonly known and widely used regularization functions.
 The former yields the popular Online Descent Gradient  (ODG)  algorithm, while the latter results in the widely used Exponential Weights Algorithm (EWA). Furthermore, \citet[p.101]{Hazan2017} notices the following: ``\emph{There are surprisingly few cases of interest besides the Euclidean and Entropic regularizations and their matrix analogues}''. 
 \smallskip

In this section, we show that a  ``new''  regularization function is available in the case of the NL model. As we discussed in \S\ref{s33}, for the NL case the generator function $G$ is given by $G(e^{\boldsymbol{\theta}_t/\eta})=\sum_{k=1}^K\left(\sum_{i\in\mathcal{N}_k}e^{\boldsymbol{\theta}_{it}/\eta\lambda_k}\right)^{\lambda_k}$ and $\varphi(\boldsymbol{\theta}_t)=\eta \log G(e^{\boldsymbol{\theta}_t/\eta})+\eta\gamma$. Exploiting this specific functional form, $\R(\bold{x})$ can be expressed in closed form.

\begin{lemma}\label{NL_regul_lemma}In the NL model the following hold:
	\begin{eqnarray}\label{NL_regularization}
	\mathcal{R}(\bold{x}_t)&=&\eta\sum_{k=1}^K\sum_{i\in\mathcal{N}_k}\lambda_k\bold{x}_{it}\log\bold{x}_{it}+\\
	&&\eta\sum_{k=1}^K(1-\lambda_k)\left(\sum_{i\in \mathcal{N}_k}\bold{x}_{it}\right)\log\left(\sum_{i\in \mathcal{N}_k}\bold{x}_{it}\right)\mbox{for $t=1,\ldots,T$.}\nonumber
	\end{eqnarray}
	
\end{lemma}

The previous lemma provides a simple expression for $\mathcal{R}(\bold{x})$, which is a generalization of the Shannon entropic term. To see this, we note that in Eq. (\ref{NL_regularization}) the first term captures the Shannon
entropy \emph{within} nests, whereas the second term captures the information \emph{between} nests. Accordingly,  $\mathcal{R}(\bold{x})$ can interpreted as  an \emph{augmented} (or generalized) version of Shannon entropy. In the context rational  inattention models, the function (\ref{NL_regularization}) was introduced by \citet{Fosgerauetal2019a}. Our proof is a simple adaptation of their arguments. However, to our knowledge,  regularization (\ref{NL_regularization}) is new to the  FTRL algorithm literature.
\smallskip

Based on Lemma \ref{NL_regul_lemma}, in the following proposition, we show that in the case of the NL model, the choice probability vector $\bold{x}_t$ has a recursive structure. 
\begin{proposition}\label{Exp_Nested1} Supposed that the FTRL algorithm is implemented with the NL model. Then the choice probability vector $\bold{x}_t$ satisfies:
	\begin{equation}\label{nested_logit_choice}
	\bold{x}_{it+1}= \left(\Phi_{i}(\bold{x}_t)\right)^{1\over \lambda_k}\PP_{ikt}\PP_{kt}\quad\forall i,k,t,
	\end{equation}
	where:
	\begin{eqnarray}
	\Phi_{i}(\bold{x}_t)&\triangleq&\bold{x}_{it}^{\lambda_k}\left(\sum_{j\in\mathcal{N}_k}\bold{x}_{jt}\right)^{1-\lambda_k},\nonumber\\
	\PP_{ikt}&\triangleq&\frac{e^{\bold{u}_{it} / \eta\lambda_k}}{\sum_{j \in \mathcal{N}_k}\Phi_{j}(\bold{x}_t) e^{\bold{u}_{jt} / \eta\lambda_k}} ,\quad\mbox{and}\nonumber\\
\PP_{kt}	&\triangleq&\triangleq \frac{ \left(\sum_{j \in \mathcal{N}_k} \Phi_{j}(\bold{x}_t)e^{\bold{u}_{jt}/ \eta\lambda_k}\right)^{\lambda_k}}{\sum_{l=1}^K  \left(\sum_{j^\prime \in \mathcal{N}_{l}} \Phi_{j^\prime}(\bold{x}_t)e^{\bold{u}_{j^\prime t} /\eta \lambda_{l}}\right)^{\lambda_l}}.\nonumber
	\end{eqnarray}

\end{proposition}

Formally, this result allows one to understand how  $\bold{x}_t$ evolves as the DM learns about past realizations of the payoff vector $\bold{u}_t$.  
\smallskip

More importantly,  Proposition \ref{Exp_Nested1}  makes explicit the recursive structure of the FTRL when we use the NL model. In particular, our result is a generalization of the EWA approach. When the values of the nesting parameters converge to one, expression (\ref{nested_logit_choice}) boils down to the recursive MNL model. The following corollary formalizes this observation.
\begin{corollary}\label{Exp_Gumbel} In Proposition \ref{Exp_Nested1} let $\lambda_k=1$ for all $k.$ Then 
	\begin{equation}\label{logit_cor}
	\bold{x}_{it+1}={\bold{x}_{it}e^{\bold{u}_{it}/\eta}\over\sum_{j=1}^N \bold{x}_{jt}e^{\bold{u}_{it}/\eta}}\quad \forall i,t.
	\end{equation}
\end{corollary}

\subsection{Recency bias}\label{Recency_Bias} So far in our regret analysis, we have assumed that the DM  weights all past observations similarly. However, there is plenty of evidence that in repeated choice problems, a DM reacts more heavily to recent observations than she does to old ones. In the learning literature this phenomenon is referred as \emph{recency bias} (e.g. \citet{ErevHaruv2016}, \citet{FudenbergPeysakhovich2014}, and \citet{FudenbergLevine2014}). 
\smallskip

In this section, we show how to incorporate the recency bias effect into the RUM-ODP model in a simple way. Following \citet{RakhlinSridharan2013} we introduce a sequence of functions $\boldsymbol{\beta}_t:\mathcal{U}^{t-1}\mapsto\mathcal{U}$  for each $t=1,\ldots, T$, which define a \emph{predictable} sequence
\begin{equation}\label{Predictable_Sequence}
 \boldsymbol{\beta}_1(\boldsymbol{0}),\boldsymbol{\beta}_{2}(\bold{u}_1),\ldots, \boldsymbol{\beta}_T(\bold{u}_1,\ldots,\bold{u}_{T-1}).
\end{equation}
Intuitively, the sequence (\ref{Predictable_Sequence}) can be seen as a way of incorporating prior knowledge about the sequence $\bold{u}_1,\ldots,\bold{u}_T$. In particular,  the sequence of functions (\ref{Predictable_Sequence}) allows us to model the magnitude and effect of recency bias by specifying different summary statistics. Accordingly, we introduce the recency bias by incorporating the sequence (\ref{Predictable_Sequence}) to modify the FTRL algorithm  as follows \footnote{For ease of notation, we  denote $\boldsymbol{\beta}_t(\bold{u}_1,\ldots,\bold{u}_{t-1})\triangleq\boldsymbol{\beta}_t$ for $t=1,\ldots,T.$}
\begin{equation}\label{OFTRL_Eq1}
\bold{x}_{t+1}=\begin{cases} \arg\max_{\bold{x}\in \Delta_N}\{-\mathcal{R}(\bold{x})\} &\mbox{for } t = 0 \\ 
\arg\max_{\bold{x}\in \Delta_N}\{\langle\boldsymbol{\theta}_t+\boldsymbol{\beta}_t,\bold{x}\rangle-\mathcal{R}(\bold{x})\} & \mbox{for } t=1,\ldots, T. \end{cases}.
\end{equation}

Expression  (\ref{OFTRL_Eq1}) makes explicit the fact that by adding the term $\boldsymbol{\beta}_t$, we obtain a variant of the  FTRL algorithm that incorporates recency bias. Following \citet{RakhlinSridharan2013}, we denote the resulting algorithm as  \emph{Optimistic} FTRL (OFTRL). 
\smallskip

Following \citet{RakhlinSridharan2013} and \citet{NIPS2015_5763}, we define three types of recency biases we use in our regret analysis.
\begin{definition}\label{Def_Recency_Bias} In the OFTRL algorithm we say that the DM exhibits:
	\begin{itemize}
		\item[a)]One step recency bias if $\boldsymbol{\beta}_t=\bold{u}_{t-1}$ for $t=1,\ldots, T.$
		\item[b)] $S$-step recency bias  if $\boldsymbol{\beta}_t={1\over S}\sum_{\tau=t-S}^{t-1}\bold{u}_\tau$ for $t=1,\ldots, T$.
		\item[c)] Geometrically discounted recency bias if $\boldsymbol{\beta}_t=\frac{1}{\sum_{\tau=0}^{t-1} \delta^{-\tau}} \sum_{\tau=0}^{t-1} \delta^{-\tau} \mathbf{u}_{\tau}$ where $\delta\in(0,1)$ is a discount factor for $t=1,\ldots, T.$
	\end{itemize}
	
\end{definition}

In the previous definition, each functional form for $\boldsymbol{\beta}_t$ captures different ways of using the information in more recent observations. As the following proposition shows, under these three types of recency bias, the OFTRL algorithm is Hannan consistent.
\begin{proposition}\label{OFTRL_Regret}
	 Let Assumptions \ref{Shocks_Assumption} and \ref{Gradient_LL} hold. In addition, assume that the following condition holds
	 \begin{equation}\label{Bound_Dual}
	 \|\bold{u}_t-\bold{u}_{t-1}\|_\ast\leq B\quad\mbox{for  $t=1,\ldots,T$.}
	 \end{equation}
	  Then the following  statements hold:
	  \begin{itemize}
	  	\item[i)] In the one-step recency bias with $\eta=\sqrt{\frac{LTB^2}{2\varphi(\bold{0})}}$ the OFTRL algorithm is Hannan consistent with:
	  	$$\textsc{R}_{OFTRL}^T\leq B\sqrt{2LT\varphi(\bold{0})}. $$
	  		  \item[ii)] In the S-step recency bias with $\eta=\sqrt{\frac{LTS^2B^2}{2\varphi(\bold{0})}}$ the OFTRL algorithm is Hannan consistent with:
	  		  $$\textsc{R}_{OFTRL}^T\leq SB\sqrt{2LT\varphi(\bold{0})}.$$
	  \item[iii)] In the geometrically discounted recency bias with $\eta=\sqrt{\frac{LTB^2}{2(1-\delta)^3\varphi(\bold{0})}}$,  the OFTRL algorithm is Hannan consistent with:
	  $$\textsc{R}_{OFTRL}^T\leq B\sqrt{\frac{2LT\varphi(\bold{0})}{(1-\delta)^3}}.$$
	  \end{itemize}
\end{proposition}

To understand the intuition  behind Proposition \ref{OFTRL_Regret}, we note that  Lemma \ref{Bound_OFTRL}  in Appendix \ref{Proofs} establishes that the regret associated with the OFRTL  algorithm satisfies the following bound:
\begin{equation}\label{OFRTL_bound_general}
	\textsc{R}_{OFTRL}^T\leq \eta\varphi(\bold{0})+{L\over 2\eta}\sum_{t=1}^T\|\bold{u}_t-\boldsymbol{\beta}_t\|_*^2.
\end{equation}

Intuitively, Eq. (\ref{OFRTL_bound_general}) establishes  that when the sequence $(\boldsymbol{\beta}_t)_{t\geq 1}$
predicts $\bold{u}_t$ well then the RUM-ODP model can  achieve low regret. Thus, when the DM exhibits recency bias, she can achieve low regret by implementing the OFTRL algorithm. 
\smallskip

Our second observation is related to the technical details in proving Proposition \ref{OFTRL_Regret}. Formally, our proof is based in adapting  the arguments in  \citet[Lemma 2]{RakhlinSridharan2013} and  \citet[Lemmas 21 and 22]{NIPS2015_5763}.  However, a fundamental difference between Proposition \ref{OFTRL_Regret} and their results is our use of the convex structure of the RUM. Thus, Proposition \ref{OFTRL_Regret} provides an economic justification for the OFTRL algorithm.
\smallskip

Finally, we remark that by using the equivalence in Theorem \ref{Equiv_SSA_FTRL}, we can derive an SSA version incorporating the recency bias effect.   In doing so,  we note that under recency bias,  the social surplus function  is
$$\varphi(\boldsymbol{\theta}_t+\boldsymbol{\beta}_{t+1})=\EE\left(\max_{i\in A}\{\boldsymbol{\theta}_{it}+\boldsymbol{\beta}_{it+1}+\epsilon_{it+1}\}\right).$$
\smallskip 

Then, using Theorem \ref{Equiv_SSA_FTRL},  we know  that $$\bold{x}_{t+1}=\nabla\varphi(\boldsymbol{\theta}_t+\boldsymbol{\beta}_{t+1})=\arg\max_{\bold{x}\in \Delta_N}\{\langle\boldsymbol{\theta}_t+\boldsymbol{\beta}_{t+1},\bold{x}\rangle-\mathcal{R}(\bold{x})\}.$$

Thus, we can naturally define an optimistic SSA which connects the notion of recency bias, the social surplus function, and the class of  RUMs.
\section{No-regret learning in games}\label{S5_Games}
In this section, we apply the RUM-ODP model to the study of no-regret learning in games. We consider a static game $\G$ among a set $\P$ of $P$ players.\footnote{In this section, we  closely follows   the exposition in  \cite{NIPS2015_5763}.} Each player $j$ has a strategy space $S_{j}$ and a utility function $\textbf{u}_{j}: S_{1} \times \ldots \times S_{P} \rightarrow[0,1]$ that maps a strategy profile $s=\left(s_{1}, \ldots, s_{P}\right)$ to a utility $u_{j}(\mathbf{s})$. We assume that the strategy space of each player is finite and has cardinality $N$, i.e. $\left|S_{i}\right|=N$. We denote with $\mathbf{x}=\left(\mathbf{x}_{1}, \ldots, \mathbf{x}_{P}\right)$ a profile of mixed strategies, where $\mathbf{x}_{j} \in \Delta_N$ and $\textbf{x}_{jk}$ is the probability of strategy $k\in S_{i}$.\footnote{We note that in this strategic environment,  $\Delta(S_j)\triangleq \Delta_N$.} The set of profile of mixed strategies  is denoted as $\Delta_N^P\triangleq\prod_{j\in \P}\Delta_{N}$ Finally let $U_{j}(\mathbf{x})=\mathbb{E}_{\mathbf{s} \sim \mathbf{x}}\left[\textbf{u}_{j}(\mathbf{s})\right]$, the expected utility of player $j$.
\smallskip

We consider a situation where the game $\G$ is played repeatedly for $T$ time steps. We denote this repeated as $\G^T$. At each time step $t$ each player $j$ chooses  a mixed strategy $\mathbf{x}_{j}^{t} \in \Delta_N$. At the end of the iteration, each player $j$ observes the expected utility he would have received had he played any possible strategy $k \in S_{j}$. More formally, let $\textbf{u}_{j k}^{t}=\mathbb{E}_{\mathbf{s}_{-j} \sim \mathbf{x}_{-j}^{t}}\left[\textbf{u}_{j}\left(k, \mathbf{s}_{-j}\right)\right]$, where $\mathbf{s}_{-j}$ is the set of strategies of all but the $j^{t h}$ player, and let $\mathbf{u}_{j}^{t}=\left(\textbf{u}_{jk}^{t}\right)_{k \in S_{j}}$. At the end of each iteration, each player $j$ observes $\mathbf{u}_{j}^{t}$. It follows that  the expected utility of a player at iteration $t$ is given by the inner product $\left\langle\mathbf{x}_{j}^{t}, \mathbf{u}_{j}^{t}\right\rangle$.
\smallskip

To model no-regret learning, we  assume that each player
decides her mixed strategy $\mathbf{x}_{i}^{t+1}$ using the SSA.  In doing so, we define player $j$'s social surplus function as  $\varphi_j(\boldsymbol{\theta}_j^t)\triangleq\EE(\max_{l\in S_j}\{\boldsymbol{\theta}_{jl}^t+\eta_j\epsilon^{t+1}_{jl}\})$, where the expectation is taken with respect to $\epsilon^{t+1}_j$ and $\eta_j>0$ is the player-specific learning parameter.
\smallskip

In this strategic setting, the SSA is defined as   $\mathbf{x}_{j}^{0}=\nabla\varphi_j(\bold{0})$ and for $t=1,\ldots,T$ we get:
$$\mathbf{x}_{j}^{t+1}=\nabla\varphi_{j}(\boldsymbol{\theta}^{t}_{j})\quad\forall j\in \P,$$
where  $\boldsymbol{\theta}^t_{j}\triangleq \sum_{l=1}^t\bold{u}^l_{j}$.
\smallskip

In the repeated game $\G^T$,  the regret  after $T$ periods  is equal to the \emph{maximum} gain that player $j \in \P$ could have achieved by switching to any other fixed strategy in hindsight:
$$\textsc{R}_{SSA_j}^T\triangleq\max _{\mathbf{x}_{j}^{*} \in \Delta_N} \sum_{t=1}^{T}\left\langle\mathbf{x}_{j}^{*}-\mathbf{x}_{j}^{t}, \mathbf{u}_{j}^{t}\right\rangle.$$
\smallskip

It is straightforward to show that under  Assumptions \ref{Shocks_Assumption} and \ref{Gradient_LL}, we can apply  Theorem \ref{SS_Surplus_algorithm_regret} to bound $\textsc{R}_{SSA_j}^T$. In particular,  in the repeated  game $\G^T,$ setting  $\eta_j=\sqrt{{L_jT\over 2\varphi_j(\bold{0})}}$, we obtain\footnote{We note that in setting  $\eta_j=\sqrt{{L_T\over 2\varphi_j(\bold{0})}}$ we have used the fact that $u_{max}=1$ for all player $j\in \P.$ In addition,  the parameter $L_j$ corresponds to the Lipschitz constant associated to  player $j$'s social surplus function $\varphi_j$.}
\begin{equation*}
	\textsc{R}_{SSA_j}^T\leq \sqrt{2\varphi_j(\bold{0})L_jT}\quad\mbox{for all $j\in \P$}.
\end{equation*}	

The main implication of using the RUM-ODP model combined with the $SSA$  is that we can bound the regret associated with each player using a large class of discrete choice models. For instance, we can consider cases where some players may use an MNL model, and others can use GNL. In general, our approach is flexible enough to accommodate players using different discrete choice models to compute $\nabla\varphi_j(\boldsymbol{\theta}^{t}_j)=\textbf{x}_{j}^{t+1}$ for each player $j\in\P.$
\smallskip

It is worth mentioning that a significant advantage of using the SSA to study no-regret learning in normal form games is that we do not need to specify the regularization term $\mathcal{R}_j(\bold{x}_j^t)$. This feature is different from most of the literature on no-regret dynamics, which focuses on the idea of regularized learning.\footnote{See, for instance, \cite{NIPS2015_5763} and \cite{Sandholm_Mertikopoulos_2016}.} In particular, our SSA approach applies even in situations where $\mathcal{R}_{j}(\bold{x}_j^t)$ may not have a closed form expression. In addition, we mention that from using the results in \S \ref{Recency_Bias}, we can combine the SSA with the notion of recency bias. Thus, our analysis can accommodate this type of learning behavior. 
\subsection{ Coarse Correlated Equilibrium}An important implication of using the RUM-ODP model to study no-regret learning in games is that we can expand the class of choice models that allow us to approximate coarse correlated equilibrium (CCE). 
Formally, a CCE is defined as follows:
\begin{definition}[Coarse Correlated Equilibrium (CCE)]
	A distribution $\sigma$ on the set $S_{1} \times \cdots \times S_{P}$ of outcomes of  the  game  $\G$ is a coarse correlated equilibrium $(C C E)$ if for every agent $i \in\P=\{1,2, \ldots, P\}$ and every unilateral deviation $s_{j}^{\prime} \in S_{j}$,
	\begin{equation}\label{CCE_Condition}
		\mathbb{E}_{\mathbf{s} \sim \sigma}\left[\bold{u}_{j}(\mathbf{s})\right] \geq \mathbb{E}_{\mathbf{s} \sim \sigma}\left[\bold{u}_{j}\left(s_{j}{ }^{\prime}, \mathbf{s}_{-j}\right)\right] .
	\end{equation}
\end{definition}
The condition (\ref{CCE_Condition}) is the same as that for a mixed strategy Nash equilibrium, except without the restriction that $\sigma$ is a product distribution.   Intuitively,  this condition applies to a situation where an agent $i$ contemplating  a deviation $s_{i}{ }^{\prime}$ knows only the distribution $\sigma$ and not the component $s_{i}$ of the realization.  In other words, a CCE only protects against \emph{unconditional unilateral deviations}, as opposed to the unilateral deviations conditioned on $s_{i}$ that are addressed in the definition of a correlated equilibrium (\cite{Aumann197467}). Furthermore,  it follows that every CE is a CCE, so a CCE is guaranteed to exist and is computationally tractable.\footnote{For an in-depth discussion of this issue we refer the reader to  \cite{Roughgarden2016}.}  More importantly,  it is well-known that no-regret dynamics converge to the set of coarse correlated equilibria (\cite[Prop. 17.9]{Roughgarden2016}). The following result adapts this observation to the case of the RUM-ODP model.
\begin{proposition}\label{CCE_Result} Let Assumptions  \ref{Shocks_Assumption} and \ref{Gradient_LL} hold.  Suppose that at periods $t=1,\ldots, T$,  players choose their strategy $\bold{x}_j^t$ according to the SSA. Let $\sigma^{t}=\prod_{j=1}^{P} \bold{x}_{j}^{t}$ denote the outcome distribution at iteration $t$ and $\sigma=\frac{1}{T} \sum_{t=1}^{T} \sigma^{t}$ the time-averaged history of these distributions. Then $\sigma$ is an approximate coarse correlated equilibrium, in the sense that
	$$
	\mathbb{E}_{\mathbf{s} \sim \sigma}\left[\bold{u}_{j}(\mathbf{s})\right] \geq \mathbb{E}_{\mathbf{s} \sim \sigma}\left[\bold{u}_{j}\left(s_{j}^{\prime}, \mathbf{s}_{-j}\right)\right]-\delta.
	$$
	for every agent $i$ and unilateral deviation $s_{i}^{\prime}$ where $\delta\triangleq \max_{j\in \P}\{\textsc{R}_{SSA_j}^T\}$.
\end{proposition}

Some remarks are in order. First, the result in Proposition \ref{CCE_Result} is well-known in the algorithmic game theory literature.\footnote{ \cite[pp. 240-241 ]{Roughgarden2016} provides a proof of this result. For completeness, we provide the proof of this result in Appendix \ref{Proofs}.}  The main innovation on it is the definition of  $\delta$ in terms of the players' regret bounds. This parameter incorporates the role of the RUM under consideration through the social surplus function $\varphi$ and the Lipschitz constant $L$. Thus, our version of this result provides a connection between no-regret dynamics, RUMs, and the set of coarse correlated equilibria of the game $\G$. 
\smallskip

Second, we point out that given the flexibility of the SSA approach,   a large class of RUMs is available to study no-regret dynamics and CCE points. This feature contrasts with most of the literature on no-regret learning, which primarily focuses on the FTRL using well-known regularization functions. In fact, to the best of our knowledge,  most of the literature on no-regret learning has concentrated on the MNL model for which the term $\mathcal{R}_j(\bold{x}_j^t)$  corresponds to the Shannon entropy. Thus, our result allows us to study no-regret learning and convergence to the set of  CCE in normal form games considering a large class of RUMs. 

\smallskip
As discussed in \S \ref{s2}, the social surplus function corresponds to a potential function. This feature implies that the SSA approach is related to  the framework of  potential-based dynamics developed in \cite{BianchiLugosi2003}, \cite{HART_JET_200126}, and \cite{HART_GEB_2003375}. However, we pointed out in \S \ref{s2} that our framework and theirs differ in at least three important aspects. First,   \cite{BianchiLugosi2003}'s approach requires the existence of a potential function that satisfies such a requirement. Our use of the social surplus function $\varphi$ (as a potential)   does not require this condition. As we discussed earlier, this is equivalent to assuming that the random preference shocks are i.i.d., which is very restrictive, ruling out a large class of RUMs. Second, the approach in \cite{HART_JET_200126} and \cite{HART_GEB_2003375} requires the existence of a potential function $P:\RR^N\mapsto\RR_+$ with the  property that the gradient $\nabla P$ vanishes over the approachable set $\RR_{-}^N$.  In terms of the RUM-ODP model, this is equivalent to assuming that the random preference shock $\epsilon$ has bounded support. This condition rules out the whole family of RUMs discussed in this paper. Finally, a third difference is related to the economic interpretation of the potential function. As we showed in Eq.  (\ref{Potential_Econ}), the social surplus function $\varphi$ has a   clear economic interpretation, which highlights the role of the cumulative payoff vector $\boldsymbol{\theta}_{j}^t$ and the random preference shock $\epsilon$. Thus, while related, our approach is different from theirs.  
\subsection{Efficiency} An important application of no-regret learning is the possibility of analyzing the average welfare in a repeated game where players are no-regret learners.  In order to formalize this,   for a given strategy profile $\textbf{s}$ the social welfare is defined as the sum of the player utilities: $$W(\mathbf{s})\triangleq\sum_{j \in \P} \textbf{u}_{i}(\mathbf{s}).$$
\smallskip

Similarly, given a mixed strategy profile $\bold{x}$, we define
$$W(\mathbf{x})\triangleq\mathbb{E}_{\mathbf{s} \sim \mathbf{x}}[W(\mathbf{s})]$$
\smallskip

Our basic goal is to set a lower bound on how far the sequence's average welfare concerns the static game's optimal welfare $\G$. Formally, we focus on the following measure:
$$\mathrm{OPT}\triangleq\max _{\mathbf{s} \in S_{1} \times \ldots \times S_{n}} W(\mathbf{s}).$$
From an economic standpoint, the optimal welfare $\mathrm{OPT}$  corresponds to a situation where players do not have incentives to be strategic and if a central planner could enforce (or dictate) each player's strategy.  Following the Algorithmic Game Theory literature, we define the class of \emph{smooth games}(\cite{Roughgarden_2015}).
\begin{definition} [\cite{Roughgarden_2015}]\label{smooth_games} A game is $(\lambda, \mu)$-smooth if there exists a strategy profile $\mathbf{s}^{*}$ such that for any strategy profile $\mathbf{s}: \sum_{j \in \P} \mathbf{u}_{j}\left(s_{j}^{*}, \mathbf{s}_{-j}\right) \geq \lambda \mathrm{OPT}-\mu W(\mathbf{s})$.
\end{definition}

Intuitively, Definition \ref{smooth_games} establishes that any player using her optimal strategy continues to do well irrespective of other players' strategies.  This condition implies \emph{near-optimality} of no-regret dynamics when we apply the RUM-ODP model.

\begin{proposition}\label{smooth_games_result} Let Assumptions  \ref{Shocks_Assumption} and \ref{Gradient_LL} hold. Then in a $(\lambda, \mu)$-smooth game, the following holds:
	$$
	\frac{1}{T} \sum_{t=1}^{T} W\left(\mathbf{x}^{t}\right) \geq \frac{\lambda}{1+\mu} \mathrm{OPT}-\frac{1}{1+\mu} \frac{1}{T} \sum_{j \in \P} \textsc{R}_{SSA_j}^T=\frac{1}{\rho} \mathrm{OPT}-\frac{1}{1+\mu} \frac{1}{T} \sum_{j \in \P} \textsc{R}_{SSA_j}^T,
	$$
	where the factor $\rho=(1+\mu) / \lambda$ is called the price of anarchy (\textsc{PoA}).
\end{proposition}
The proof of Proposition \ref{smooth_games_result} follows the arguments  in \cite[Prop. 2]{NIPS2015_5763} and \cite{Roughgarden_2015}.  We contribute to this result by providing an explicit expression for the terms $\textsc{R}_{SSA_j}^T$. In particular, given that the \textsc{PoA} is driven by  the quantity $\frac{1}{1+\mu} \frac{1}{T} \sum_{j \in \P} \textsc{R}_{SSA_j}^T$, we can  connect the no-regret learning behavior with a large class of RUMs. In particular,   given Assumptions \ref{Shocks_Assumption} and \ref{Gradient_LL}, we know that  $\textsc{R}_{SSA_j}^T=O(\sqrt{\varphi_j(\bold{0})L_jT})$. From   \S \ref{s2} we know that this  explicit bound captures the structure of the RUM-ODP model through the parameters $\varphi_j(\textbf{0})$ and $L_j$.   Thus, our contribution is to provide economic content to the fact that by employing the SSA, the average welfare converges to the \textsc{PoA}. 
\section{Prediction markets and the RUM-ODP model}\label{s555}
A prediction market is a future market in which prices aggregate information and predict future events (\citet{Hanson02logarithmicmarket}). The designer of these markets is called a market maker, and her goal is to incentivize accurate predictions of uncertain outcomes. In these markets, goods correspond to securities with payoffs contingent on uncertain outcomes. Applications of prediction markets include electoral markets, science and technology events, sports events, the success of movies, etc. (\citet{WolfersZitzewitz2004}).
\smallskip

Recently, \citet{ChenWortman2010} and \citet{Abernethyetal2014} have established the connection between prediction markets and online learning models. They show that a general class of cost-function-based prediction markets corresponds to applying the FTRL algorithm to these specific markets.
\smallskip

This section shows how the RUM-ODP  and the SSA can be used to study prediction markets. In doing so, we exploit the mathematical structure of the social surplus function, which enables us to connect the SSA with a large class of prediction markets. In economic terms, our analysis establishes a formal relationship between machine learning, the RUM-ODP  model, and prediction markets.
 \subsection{Cost-function based prediction markets}\footnote{This section closely follows the discussion of cost-function-based markets described in \citet{ChenWortman2010}.} A popular approach in the study of prediction markets is the idea of cost-function-based prediction markets. In this setting, there is an agent denoted as the market maker, who trades a set of securities corresponding to each potential outcome of an event. Formally, let   $\Omega=\{1, \cdots, N\}$ be a set of \emph{mutually} exclusive and exhaustive outcomes of a particular event (for instance, an electoral result or the winner in a sporting event). To price the securities associated with the uncertain events, the market maker uses a differentiable cost function $C: \mathbb{R}^{N} \rightarrow \mathbb{R}$ to determine the prices. This cost function describes the amount of money currently wagered in the market as the number of shares purchased.
 Let  $\bold{q}_{i}$ be the  number of shares of security $i$ currently held by traders for $i=1,\ldots,N$. Accordingly, let $\bold{q}=(\bold{q}_1,\ldots,\bold{q}_N)$ be the vector of shares. A trader would like to purchase $\bold{r}_{i}$ shares of each security paying $C(\bold{q}+\bold{r})-C(\bold{q})$ to the market maker, where $\bold{r}=(\bold{r}_1,\ldots,\bold{r}_N)\in \RR^N$.\footnote{The share $\bold{r}_{i}$ could be zero or even negative, representing the sale of shares. Thus the vector $\bold{r}$ can be treated as an element of $\RR^N$.} 
 \smallskip

Given an infinitely small number of shares, the instantaneous price of security $i$ is given by $p_{i}(\bold{q})={\partial C(\bold{q}) \over \partial \bold{q}_{i}}$.  In other words,  the price per share of an infinitely small number of shares is given by 
the gradient of the cost function $C$.

\citet{ChenWortman2010} define a cost function $C$ to be valid  if the associated prices
satisfy the following  two simple conditions:
\begin{enumerate}
	\item For every  $i \in 1, \ldots, N$ and every $\bold{q} \in \mathbb{R}^{N}, p_{i}(\bold{q}) \geq 0$.
	\item For every $\bold{q}\in \mathbb{R}^{N}, \sum_{i=1}^{N} p_{i}(\bold{q})=1$.
\end{enumerate}
Condition 1 establishes that the price of a security is non-negative. In the case of a negative price $p_i(\bold{q})$, a trader could purchase shares of this security at a guaranteed profit. Similarly, condition 2  establishes that the prices of all securities add up to 1. If it were the case that $\sum_{i=1}^{N} p_{i}(\bold{q})<1$, then a trader could purchase  small equal quantities of each security for a guaranteed profit.\footnote{In the case  $\sum_{i=1}^{N} p_{i}(\bold{q})>1$, then a trader could sell  small equal quantities of each security for a guaranteed profit. } Combining these two requirements on prices, we ensure that there are no arbitrage opportunities within the market. More importantly, conditions 1 and 2 allow us to interpret the price vector as a valid probability distribution over the outcome space. In particular, these prices represent the market's current estimate of the probability that outcome $i$ will occur (\citet{MANSKI2006425}). \citet{ChenWortman2010} provide necessary and sufficient conditions for the cost function $C$ to be valid. Given the relevance to our analysis, we state  their result  for completeness
\begin{theorem}\label{ChenWortmanThm}[Chen and Vaughan (2010)]A cost function $C$ is valid if and only if it satisfies the following three properties:
	\begin{enumerate}
		\item[1.] \textsc{Differentiability:}  The partial derivatives ${\partial C(\bold{q}) \over \partial \bold{q}_{i}}$ exist for all $\bold{q} \in \mathbb{R}^{N}$ and $i \in\{1, \ldots, N\}$
		\item[2.] \textsc{Increasing Monotonicity:} For any $\bold{q}$ and $\bold{q}^{\prime}$, if $\bold{q}\geq \bold{q}^{\prime}$, then $C(\bold{q}) \geq C\left(\bold{q}^\prime\right)$
		\item[3.] \textsc{Positive Translation Invariance:} For any $\bold{q}$ and any constant $k, C(\bold{q}+k \bold{1})=C(\bold{q})+k$
	\end{enumerate}
	
\end{theorem}

The previous theorem characterizes a valid  cost function in terms of  three natural conditions. 
\citet{Hanson2003} and \citet{ChenPennock2007}  show that the cost function

\begin{equation}\label{Logsum_cost}
C(\bold{q})=b\log\left(\sum_{i=1}^Ne^{\bold{q}_i/b}\right)\quad \quad b>0,
\end{equation}   satisfies the requirements of Theorem \ref{ChenWortmanThm}. Moreover, taking the partial derivative with respect to $\bold{q}_i$, it is easy to see that  the pricing function $p_i(\bold{q})$  is given by:
\begin{equation}\label{Price_function}
p_i(\bold{q})={e^{\bold{q}_i/b}\over \sum_{j=1}^Ne^{\bold{q}_j/b}}, \quad \mbox{for $i=1,\ldots, N.$}
\end{equation}

In  the prediction markets literature,   expressions (\ref{Logsum_cost}) and   (\ref{Price_function}) define a \emph{Logarithmic Market Scoring Rule} (LMSR) which was introduced by \citet{Hanson2003,Hanson02logarithmicmarket}. In particular, the prices in (\ref{Price_function}) allows us to capture a situation in which a trader who changes the market probabilities from $\bold{r}$ to $\bold{r}^{\prime}$ obtains the same payoff for every outcome $i$ as a trader who changes the quantity vectors from any $\bold{q}$ to $\bold{q}^{\prime}$ such that $p(\bold{q})=\bold{r}$ and $p\left(\bold{q}^{\prime}\right)=\bold{r}^{\prime}$ in the cost function formulation (\citet{ChenWortman2010} and \citet{Abernethyetal2013}).
\smallskip

Based on our discussion in \S \ref{s33},  it is easy to see that functions (\ref{Logsum_cost}) and  (\ref{Price_function}) can be seen as a particular application of the MNL  to prediction markets. More formally, identifying $\boldsymbol{\theta}=\bold{q}$ and $\eta=b$, we can conclude that  for the MNL model, the social surplus function $\varphi(\boldsymbol{\theta})$ can be interpreted as a cost function.  Strikingly, this relationship is far more general, as the next proposition shows.
\begin{proposition}\label{SS_equiv_CBM} Let Assumption \ref{Shocks_Assumption} hold. Then the social surplus function is a valid cost function.
\end{proposition}

This result directly implies that the RUM  is useful for studying prediction markets. More importantly, Proposition \ref{SS_algorithm_GEV} allows us to implement the SSA in the context of prediction markets.   \citet{ChenWortman2010}  and \citet{Abernethyetal2013} pointed out the connection between cost-function-based markets and online learning algorithms. They show that the FTRL algorithm is useful for constructing pricing mechanisms in a dynamic environment.  Intuitively, this equivalence establishes that the DM uses the FTRL algorithm  to select a probability distribution $\bold{x}$ while the market maker uses a 
\emph{duality-based} cost function to compute the price vector $\bold{p}(\bold{q})$. Unfortunately, this connection relies on knowing the convex conjugate of the cost function $C$.
\smallskip

By combining  Theorem \ref{Equiv_SSA_FTRL} with   Proposition \ref{SS_equiv_CBM}  we can connect  the RUM-ODP and the SSA  with prediction markets. In doing so, we identify outcomes in $\Omega$ with alternatives in a discrete choice set $A$ and trades
made in the market with payoffs observed by the SSA. Thus, we can view the market maker as learning
a probability distribution over outcomes by treating each
observed trade $\bold{r}_t$ as a realization of the environment in the same fashion as the 
SSA allows the DM learns a distribution $\bold{x}_{t+1}$ over the set $A=\{1,\ldots, N\}$ using observed realizations of $\bold{u}_t$. Using this analogy, we can rewrite the SSA in terms of prediction markets:
   \begin{algorithm}
   	\caption{Prediction Market Algorithm}\label{alg:euclid1}
   	\begin{algorithmic}[1]
   		\State Input: $\eta>0$, $F$ a distribution on $\RR^N$, and $\Delta_N$.
   		\State Let $\bold{q}_0\in\RR^N$ and choose  $\bold{x}_1=\nabla\varphi(\bold{q}_0)$ 
   		\State $\bold{for}$ $t=1$ to $T$ \textbf{do}
   		\begin{itemize}
   			\item The market maker sets prices $\bold{x}_{t}=\nabla\varphi(\bold{q}_{t-1})$
   			\item The market maker receives security bundle purchase  $\bold{r}_{t}$
   			\item The market maker obtains the expected  payoff $\langle\bold{u}_{t}, \nabla\varphi(\bold{q}_{t-1})\rangle$
   			\item The market maker updates accordingly to $\bold{q}_{t}=\bold{r}_t+\bold{
   				q}_{t-1}$ and chooses
   			$$\bold{x}_{t+1}=\nabla\varphi(\bold{q}_t)$$
   		\end{itemize}
   		\State \textbf{end for}
   	\end{algorithmic}
   	\end{algorithm}

Given our results in GEV models, it is easy to see that the previous algorithm opens the possibility of using several new cost and pricing functions in the context of cost-function-based prediction markets. Moreover,  the regret analysis is similar to the arguments behind Theorem \ref{SS_Surplus_algorithm_regret}. We leave for future work a more profound analysis of the connection between prediction markets, the SSA, and the RUM-ODP model.

\section{Related literature}\label{s5}
Regret theory was introduced in a series of seminal papers by \citet{Bell1982},  \citet{LoomesSugden1982, LOOMES1987270} and  \citet{FISHBURN198231} as an alternative to the expected utility paradigm. In simple terms, regret theory establishes that a DM wants to avoid outcomes in which she will appear to have made the wrong decision, even if,  in advance, the decision appeared correct with the information available at the time. In particular, regret theory entails the possibility of non-transitive pairwise choices.\footnote{For a complete discussion of regret theory and its contributions, we refer the reader to  \citet{BleichrodtWaakker2015}. } Recently, \citet{Sarver2008} and \citet{HAYASHI2008242} provide an axiomatic foundation for regret preferences. Our paper contributes to this literature by studying algorithmically the no-regret concept in the context of the RUM-ODP model. In this sense, our regret analysis is closer to the one in the algorithmic game theory literature (\citet{Roughgarden2016}). 

\smallskip

Our paper mainly relates to the literature on no-regret dynamics in repeated games. As we mentioned earlier, \citet{Hahn1957}'s seminal work introduces the idea of \emph{consistency} as a benchmark when considering a sequence of repeated play. The papers by \citet{LittlestoneWarmuth1994}, \citet{FudenbergLevine1995}, \citet{FreundSchapire1997}, \citet{FreundSchapire1999}, \citet{BlumMansur2007}, \citet{FosterVohra1997}, and \citet{HartMascolell200}, among many others, extend Hannan's analysis to different strategic environments.\footnote{ For an in-depth analysis of no-regret learning in games, we refer the reader to  \citet{CesaBianchiLugosi2006}.} Recently, \citet{NIPS2015_5763} studied the fast convergence of online learning in the context of regularized games using the  FTRL algorithm. Our work differs from these papers in at least two aspects. First, we introduce the SSA, which allows us to study the RUM-ODP model, exploiting the theory of discrete choice models. In particular, we show how the RUM-ODP provides closed-form expressions for several discrete choice models. Second, we provide a regret analysis and a generalization of EWA not covered by the papers cited above.

\smallskip

The papers by \cite{BianchiLugosi2003}, \cite{HART_JET_200126}, and \cite{HART_GEB_2003375} study  no-regret learning in normal form games exploiting the notion of potential functions. In particular, these papers introduce the notion of potential-based learning. As discussed in the main text, our SSA approach is an instance of a potential-based algorithm.   However, our framework differs from this line of work in at least three aspects. First, we do not impose the additivity condition used by \cite{BianchiLugosi2003}. Second,  we do not impose any condition in the domain of the gradient social surplus.   \cite{HART_JET_200126} and \cite{HART_GEB_2003375} impose the condition  that over the approachable set the gradient of their potential function vanishes. This condition is incompatible with the class of RUMs, implying that their results do not apply to the RUM-ODP model.   
\smallskip

As we mentioned in \S \ref{s1}, the RUM-ODP model is an instance of a two-person game between the DM and the environment. In this sense, our paper is related to the recent work by \citet{gualdani2020identification}. They study a discrete choice model in which the DM possesses imperfect information about the utility generated by the available options. They model this problem as an incomplete information game between the DM and the environment, exploiting the notion of \emph{Bayes Correlated Equilibrium} (\citet{BergemannMorris2016}). Our paper differs from theirs in at least three crucial aspects. First, \citet{gualdani2020identification} analyze a static incomplete information game while we study a  repeated choice situation. Second, we focus on understanding under which conditions the RUM-ODP model achieves Hannan consistency,  while \citet{gualdani2020identification}'s goal is the econometric identification of the DM's preferences. Third, our analysis focuses on discrete choice models. \citet{Magnolfi2021} study no-regret learning in the context of Bayes-coarse correlated equilibrium. Their main goal is the identification and econometric estimation of the structural parameters describing the underlying game. They do not study the RUM.

\smallskip

Our paper naturally connects with the OCO literature.\footnote{For an excellent treatment of the OCO problem, we refer the reader to \citet{Shalev-Shwartz2012} and  \citet{Hazan2017}.} From this literature, the work by \citet{Abernethyetal2016} is the closest to our paper. They connect the MNL model with the FTRL approach. Our paper differs from theirs in at least three fundamental aspects. First, we show that the entire class of discrete choice models naturally defines a regularization term to implement the FTRL algorithm. Second, we identify a Lipschitz condition on the gradient of the social surplus function, which allows one to characterize the class of discrete choice models that are Hannan consistent. Third, we generalize in a non-trivial way the EWA algorithm. Concretely, we introduce the NL model providing a new closed-form regularization penalty term.
\smallskip

Our paper is also related to the active and increasing literature on stochastic choice and information frictions. The papers by \citet{CaplinDean2015}, \citet{Matejka2015}, \citet{CaplinMartin2015}, \citet{Caplinetal2019}, and \citet{Fosgerauetal2019a} study (static) stochastic choice and information acquisition under the Rational Inattention (RI) framework. These papers focus on understanding the relationship between how different cost functions determine stochastic choice behavior.  \citet{Natenzon2019} develops a Bayesian Probit approach where the DM observes a noisy signal of the utility associated with each alternative in the choice set.   A  shared feature of these papers is the assumption that the DM has a \emph{prior} over the set of possible payoff realizations. In a different framework, \citet{Lu2016}  studies a RUM in which the DM has private information before deciding. He provides results where observed stochastic choice behavior is useful to recover private information. Our paper differs from this line of work in at least three aspects. First, our approach neither specifies priors over the payoff realizations nor posterior beliefs. Second, we consider a repeated stochastic choice situation, while these papers focus on a static environment. Third, our analysis focuses on Hannan consistency as a performance benchmark,  while the above-cited papers focus on the utility maximization paradigm. 
\smallskip

The recent contributions  by \citet{WEBB2019} and \citet{cerreiavioglio2021multinomial} are the closest papers to our work.  The paper by Webb derives the  RUM  using a general class of bounded accumulation models. This connection allows him to characterize the resulting distribution of the stochastic component in a RUM based on response times. The paper by \citet{cerreiavioglio2021multinomial}  provides an axiomatic characterization of the MNL in which time-constrained information processing causes stochastic choice behavior. In addition, they propose a neural approach that provides a causal analysis of the decision maker
choices through a biologically inspired algorithmic decision process. Our paper differs  from   \citet{WEBB2019} and \citet{cerreiavioglio2021multinomial} in at least three important aspects. First, in the RUM-ODP model, the DM learns through repeated choice, while in \citet{WEBB2019} and \citet{cerreiavioglio2021multinomial}  time is used to accumulate evidence before making a choice. Second, we focus on the notion of regret while they study random utility maximization. Third, our approach is algorithmic.\footnote{To provide a neurophysiological foundation to the MNL model, \citet{cerreiavioglio2021multinomial} propose the Metropolis-DDM algorithm to model how the DM acquires information.  Our algorithm uses ideas from the machine learning and OCO literature.}
\smallskip

Our paper is also related to the literature on stochastic choice and perturbed utility models. In particular, our paper is related to the work by \citet{Fudenbergetal2015}. They show that stochastic choice corresponds to the optimal solution of maximizing the sum of expected utility and a nonlinear perturbation. This latter term is a regularization function in the language of the FTRL algorithm. While related, the work by \citet{Fudenbergetal2015}  studies perturbed utility from an axiomatic standpoint without considering learning.  
\smallskip

Finally, from a technical point of view, our paper is related to the recent contribution by \citet{Nesterovetal2019}. In particular, our results exploit their connection between the social surplus function and the concept of proxy functions. However, \citet{Nesterovetal2019} do not study the problem of no-regret learning.

\section{Final Remarks }\label{s6}
This paper proposes the RUM-ODP model to study no-regret learning in uncertain environments. Our approach is algorithmic, providing a connection between the theory of RUMs and the gradient(potential)-based learning dynamics. In particular, we introduced the SSA framework, which allows us to apply a large class of discrete choice models to analyze online decision-making and no-regret learning problems. In addition, we showed that the popular FTRL algorithm has a clear and meaningful economic interpretation. Exploiting this fact, we establish a recursive structure to the choice probability vector generated by the FTRL algorithm. This latter fact generalizes in a non-trivial way the exponential weights algorithm to discrete choice models far beyond the MNL case. In terms of applications, we use our framework to study no-regret learning in normal form games and implement prediction markets.
\smallskip

Finally, we mention that several extensions are possible. First, we relax the complete information assumption in ongoing work by studying the RUM-ODP model using a bandit approach (\citet{lattimore_szepesvari_2020}). Second, given the structure of the RUM-ODP  model, an important implication of the results derived in this paper is the possibility of studying the econometrics of no-regret learning in discrete choice models.

\bibliographystyle{plainnat}
\bibliography{References_RUM_OCO}

\newpage

\appendix
\section{Proofs}\label{Proofs}

We begin stating two technical results that will be used throughout this appendix.

\begin{lemma}[Baillon-Haddad Theorem]\label{Baillon_Haddad} The following statements are equivalent
	\begin{itemize}
		\item[i)] $h: \mathbb{E} \rightarrow \mathbb{R}$ is convex  and differentiable with  gradient $\nabla h$ which is Lipschitz continuous with
		respect to $\|\cdot\|_{\mathbb{E}}$ with constant $L>0$.
		\item[ii)]The convex conjugate $h^{*}: \mathbb{E}^{*} \rightarrow(-\infty, \infty]$ is ${1\over L}$-strongly convex with respect to  the dual norm $\|\cdot\|_{\mathbb{E}^*}^{*}$.
	\end{itemize}
\end{lemma}
\proof\citet[ Thm. 12, Section H]{RockafellarWets1997}.\footnote{We remark that in this theorem $\mathbb{E}^{*}$ denotes the dual space of $\mathbb{E}$ and $\|\cdot\|^*_{\mathbb{E}^*}$ denotes its corresponding dual norm.} \eproof
\smallskip

The next lemma establishes the differentiability of $\mathcal{R}(\bold{x}).$
\begin{lemma}\label{Convexity_smoothness}Let Assumption \ref{Shocks_Assumption}  hold. Then $\mathcal{R}$ is differentiable.
\end{lemma}\proof The proof follows from a direct application of \citet[Thm. 5]{SorensenForsgerau2020} or \citet[Thm. 2]{galichon2021cupids}.\eproof

 \subsection{Proof of Proposition \ref{SS_Equiv}} Note that  by definition  $\varphi(\boldsymbol{\theta}_t)=\EE(\tilde{\varphi}(\boldsymbol{\theta}_t+\eta\epsilon_{t+1}))$. Then combining (\ref{FTPL3}) with \citet[{Prop. 2.3}]{Bertsekas1973} it follows that 
\begin{eqnarray}
\tilde{\bold{x}}_{t+1}&\in & \partial\tilde{\varphi}(\boldsymbol{\theta}_t+\eta\epsilon_{t+1}),\nonumber\\
\EE(\tilde{\bold{x}}_{t+1})&\in &\EE(\partial\tilde{\varphi}(\boldsymbol{\theta}_{t}+\eta\epsilon_{t+1})),\nonumber\\
&=& \partial\EE(\tilde{\varphi}(\boldsymbol{\theta}_t+\eta\epsilon_{t+1})),\nonumber\\
&=&\nabla\EE(\tilde{\varphi}(\boldsymbol{\theta}_t+\eta\epsilon_{t+1}))=\nabla\varphi(\boldsymbol{\theta}_t).\nonumber
\end{eqnarray}\eproof

\subsection{Proof of Lemma \ref{Gradient_Lipschitz}} In proving this lemma we use the fact  that  $$\varphi(\boldsymbol{\theta})=\EE\left(\max_{j=1,\ldots, N}\{\boldsymbol{\theta}_{j}+\eta\epsilon_j\}\right)=\eta\EE\left(\max_{j=1,\ldots, N}\{\boldsymbol{\theta}_{j}/\eta+\epsilon_j\}\right)=\eta\varphi(\boldsymbol{\theta}/\eta).$$ 
Thus, it is easy to see that 
$\nabla\varphi(\boldsymbol{\theta})=\nabla\varphi(\boldsymbol{\theta}/\eta)$. In addition, simple algebra shows that  ${\partial^2\varphi(\boldsymbol{\theta})\over \partial \boldsymbol{\theta}_j\partial \boldsymbol{\theta}_i}={1\over \eta }{\partial^2\varphi(\boldsymbol{\theta}/\eta)\over \partial \boldsymbol{\theta}_j\partial \boldsymbol{\theta}_i},$ for all $i,j=1,\ldots, N.$ Under this equivalence we note that 
the condition in Assumption \ref{Gradient_LL} can be rewritten as $2Tr(\varphi(\boldsymbol{\theta}/\eta))\leq L$.
\smallskip

Define the function $f(t)=\nabla\varphi(\boldsymbol{\theta}_1/\eta+t(\boldsymbol{\theta}_2/\eta-\boldsymbol{\theta}_1/\eta))$  with $f^\prime(t)=\langle\nabla^2\varphi(\boldsymbol{\theta}_1/\eta+t(\boldsymbol{\theta}_2/\eta-\boldsymbol{\theta}_1/\eta)),\boldsymbol{\theta}_2/\eta-
\boldsymbol{\theta}_1/\eta\rangle$. Noticing that 
\begin{eqnarray}
\nabla\varphi(\boldsymbol{\theta}_2/\eta)-\nabla\varphi(\boldsymbol{\theta}_1/\eta)=f(1)-f(0)&=&\int_{0}^1f^\prime(t)dt\nonumber\\
&=&\int_{0}^1\nabla^2\varphi(\boldsymbol{\theta}_1/\eta+t(\boldsymbol{\theta}_2/\eta-\boldsymbol{\theta}_1/\eta))(\boldsymbol{\theta}_2/\eta-\boldsymbol{\theta}_1/\eta)dt.\nonumber\\
\|\nabla\varphi(\boldsymbol{\theta}_2/\eta)-\nabla\varphi(\boldsymbol{\theta}_1/\eta)\|_1&\leq& \int_{0}^1\|\nabla^2\varphi(\boldsymbol{\theta}_1/\eta+t(\boldsymbol{\theta}_2/\eta-\boldsymbol{\theta}_1/\eta))(\boldsymbol{\theta}_2/\eta-
\boldsymbol{\theta}_1/\eta)\|_1dt\nonumber\\
&\leq &\int_{0}^1\|\nabla^2\varphi(\boldsymbol{\theta}_1/\eta+t(\boldsymbol{\theta}_2/\eta-\boldsymbol{\theta}_1/\eta))\|_{\infty,1}\|\boldsymbol{\theta}_2/\eta-\boldsymbol{\theta}_1/\eta\|_1\nonumber
\end{eqnarray}

To complete the proof we stress two properties of the Hessian. First, each row (or columns) of $\nabla^2\varphi(\boldsymbol{\theta}_t/\eta)$ sums up to $0$. To see this we note that $\sum_{i=1}^N\nabla_i\varphi(\boldsymbol{\theta}_t/\eta)=1$. Then simple differentiation yields $\sum_{j=1}^N{1\over \eta}\nabla_{ij}\varphi(\boldsymbol{\theta}_t/\eta)=0$ for all $i=1,\ldots, N.$ Second,  it is well known that  the off-diagonal elements of ${1\over \eta}\nabla^2\varphi(\boldsymbol{\theta}_t/\eta)$ are nonnegative (\citet[Ch. 5]{mcf1}). To see why this is true, we recall that for alternative $i$ the choice probability is given by: $\nabla_i\varphi(\boldsymbol{\theta}_t/\eta)=\PP(i=\arg\max_{j\in A}\{\boldsymbol{\theta}_{jt}+\eta\epsilon_{jt}\})$. Then increasing the terms $\boldsymbol{\theta}_{jt}$ for $j\neq i$  cannot increase the probability of choosing $i$, which is formalized as ${1\over \eta}\nabla_{ij}\varphi(\boldsymbol{\theta}_t/\eta)\leq 0$.

Now, using previous observation, we have that for a convex combination  $\tilde{\boldsymbol{\theta}}/ \eta=\boldsymbol{\theta}_1/ \eta+t(\boldsymbol{\theta}_2/ \eta-\boldsymbol{\theta}_1/ \eta)$ we have
\begin{eqnarray}
{1\over \eta}\|\nabla^2\varphi(\tilde{\boldsymbol{\theta}}/\eta)\|_{\infty,1}&=&{1\over \eta}\max_{\|\bold{v}\|\leq 1}\{\|\nabla^2\varphi(\tilde{\boldsymbol{\theta}}/\eta)\bold{v}\|_1\}\nonumber\\
&\leq& {1\over \eta}\sum_{i=1}^N\sum_{j=1}^N|\nabla_{ij}^2\varphi(\tilde{\boldsymbol{\theta}})|\nonumber\\
&=& {1\over \eta}2Tr(\nabla^2\varphi(\tilde{\boldsymbol{\theta}}))\leq {L\over \eta},\nonumber
\end{eqnarray}
where the last inequality follows from Assumption \ref{Gradient_LL}. 
Plugging  in, we arrive to the conclusion
$$\|\nabla\varphi(\boldsymbol{\theta}_2/\eta)-\nabla\varphi(\boldsymbol{\theta}_1/\eta)\|_1\leq {L }\|\boldsymbol{\theta}_2/\eta-\boldsymbol{\theta}_1/\eta\|_1\quad\forall \boldsymbol{\theta}_1,\boldsymbol{\theta}_2.$$

Finally, using the fact $\nabla\varphi(\boldsymbol{\theta})=\nabla\varphi(\boldsymbol{\theta}/\eta)$
we get
$$\|\nabla\varphi(\boldsymbol{\theta}_2)-\nabla\varphi(\boldsymbol{\theta}_1)\|_1\leq {L\over \eta }\|\boldsymbol{\theta}_2-\boldsymbol{\theta}_1\|_1\quad\forall \boldsymbol{\theta}_1,\boldsymbol{\theta}_2.$$
\eproof
\begin{lemma}\label{Positive_R_Bound_D}Let Assumption \ref{Shocks_Assumption} hold. Then	$\mathcal{R}(\bold{x})\leq 0$ for all $\bold{x}\in \Delta_N$.
	
\end{lemma}
\proof  First, we note that the Fenchel equality implies $\mathcal{R}(\bold{x})=\langle \boldsymbol{\theta},
\bold{x}\rangle-\varphi(\boldsymbol{\theta})=\langle \boldsymbol{\theta},
\bold{x}\rangle-\eta\varphi(\boldsymbol{\theta}/\eta)$ with $\bold{x}=\nabla\varphi(\boldsymbol{\theta})=\nabla\varphi(\boldsymbol{\theta}/\eta)$. We recall  that $\varphi(\boldsymbol{\theta})=\EE(\max_{j=1,\ldots,n}\{\boldsymbol{\theta}_j+\eta\epsilon_j\})$. Given that  $\max\{\cdot
\}$ is a convex function, by Jensen's inequality we get $\max_{j=1,\ldots, n}\EE(\boldsymbol{\theta}_j+\eta\epsilon_j)\leq \EE(\max_{j=1,\ldots,n}\{\boldsymbol{\theta}_j+\eta\epsilon_j\}).$
Then it follows that 
\begin{eqnarray}
\mathcal{R}(\bold{x})&=& \langle \boldsymbol{\theta},\bold{x}\rangle-\varphi(\boldsymbol{\theta}),\nonumber\\
&\leq&\langle \boldsymbol{\theta},\bold{x}\rangle-\max_{j=1,\ldots, n}\EE(\boldsymbol{\theta}_j+\eta\epsilon_j)\nonumber\\
&=& \langle \boldsymbol{\theta},\bold{x}\rangle-\max_{j=1,\ldots, n}\boldsymbol{\theta}_j,\nonumber\\
\mathcal{R}(\bold{x})&\leq& 0\nonumber
\end{eqnarray}
where the last inequality follows from the fact  that $\bold{x}\in \Delta_N$.\eproof

\begin{lemma}\label{Bregman_Bound}Let Assumptions \ref{Shocks_Assumption} and \ref{Gradient_LL} hold. Then
	$$D_{\varphi}(\boldsymbol{\theta}_t||\boldsymbol{\theta}_{t-1})\leq {L\over 2\eta}u^2_{max}.$$
\end{lemma}
\proof 
Using a second order Taylor expansion of $\varphi(\boldsymbol{\theta}_t)$ we get:

\begin{eqnarray}
\varphi(\boldsymbol{\theta}_{t+1})&=&\varphi(\boldsymbol{\theta}_t)+\langle\nabla\varphi(\boldsymbol{\tilde{\theta}}),\bold{u}_t\rangle+{1\over 2}\langle\bold{u}_t,\nabla^2\varphi(\tilde{\boldsymbol{\theta}})\bold{u}_t\rangle,\nonumber\\
\varphi(\boldsymbol{\theta}_{t+1})-\varphi(\boldsymbol{\theta}_t)-\langle\nabla\varphi(\boldsymbol{\tilde{\theta}}),\bold{u}_t\rangle&=&{1\over 2}\langle\bold{u}_t,\nabla^2\varphi(\tilde{\boldsymbol{\theta}})\bold{u}_t\rangle,\nonumber\\
D_\varphi(\boldsymbol{\theta}_{t+1}||\boldsymbol{\theta}_t)&=&{1\over 2}\langle\bold{u}_t,\nabla^2\varphi(\tilde{\boldsymbol{\theta}})\bold{u}_t\rangle,\label{Bregman_1}
\end{eqnarray}
where $\tilde{\boldsymbol{\theta}}$ is some convex combination of $\boldsymbol{\theta}_{t+1}$ and $\boldsymbol{\theta}_t$. From Eq.(\ref{Bregman_1})  it follows that
\begin{equation}\label{Bregman2}
D_\varphi(\boldsymbol{\theta}_{t+1}||\boldsymbol{\theta}_t)\leq {1\over 2}\|\nabla_{ij}^2\varphi(\tilde{\boldsymbol{\theta}})\|_{\infty,1}\|\bold{u}_t\|_{\infty}^2.\end{equation}

Noting that $$\|\nabla^2\varphi(\tilde{\boldsymbol{\theta}})\|_{\infty,1}=\max_{\|\bold{v}\|\leq 1}\{\|\nabla^2\varphi(\tilde{\boldsymbol{\theta}})\bold{v}\|_1\}\leq \sum_{i=1}^N\sum_{j=1}^N|\nabla_{ij}^2\varphi(\tilde{\boldsymbol{\theta}})|=2Tr(\nabla^2\varphi(\tilde{\boldsymbol{\theta}}))\leq {L\over \eta},$$
where the last inequality follows from Assumption \ref{Gradient_LL}.

 Plugging the previous bound in (\ref{Bregman2}) combined with $\|\bold{u}_t\|^2_{\infty}\leq u_{max}^2$    we find that                                                                                                                                                                                                                                                                                                                                                                                                                                                                                                                                                                                                                                                                                                                                                                                                                                                                                                                                                                                                                                                                                                                                                                                                                                                                                                                                                                                                                                                                                                                                                                                                                                                                                                                                                                                                                                                                                                                                                                                                                                                                                                                                                                                                                                                                                                                                                                                                                                                                                                                                                                                                                                                                                                                                                                                                                                                                                                                                                                                                                                                                                                                                                                                                                                                                                                                                                                              
$$D_\varphi(\boldsymbol{\theta}_{t+1}||\boldsymbol{\theta}_t)\leq{L\over 2\eta}u_{max}^2.$$\eproof

\begin{lemma}\label{Bound_Regret_FTRL}Let Assumptions \ref{Shocks_Assumption} and \ref{Gradient_LL} hold. Then in the SSA 
	\begin{equation}\label{FRTL_Bounded}
	\textsc{Regret}_T\leq \eta\varphi(\bold{0})+{L\over 2\eta}Tu^2_{max}.
	\end{equation}
\end{lemma}
\proof The proof of this lemma exploits the convex duality structure of the RUM-ODP model. 
By the Fenchel-Young inequality we know that 

$$\forall\bold{x}\in\Delta_N:\quad\mathcal{R}(\bold{x})\geq \langle\boldsymbol{\theta}_T,\bold{x}\rangle-\varphi(\boldsymbol{\theta}_T),$$
where the equality holds when $\bold{x}$ maximizes $\langle\boldsymbol{\theta}_T,\bold{x}\rangle-\mathcal{R}(\bold{x})$.

The Fenchel-Young inequality implies
$$\mathcal{R}(\bold{x})-\langle\boldsymbol{\theta}_T,\bold{x}\rangle\geq -\varphi(\boldsymbol{\theta}_T).$$

Noting that $-\varphi(\boldsymbol{\theta}_T)$ can be equivalently expressed as:
$$-\varphi(\boldsymbol{\theta}_T)=-\varphi(\boldsymbol{\bold{0}})-\sum_{t=1}^T\left(\varphi(\boldsymbol{\theta}_{t})-\varphi(\boldsymbol{\theta}_{t-1})\right).$$

From the definition of Bregman divergence combined with  $\bold{x}_t=\nabla\varphi(\boldsymbol{\theta}_{t-1})$, it follows that 
$$\sum_{t=1}^T\left(\varphi(\boldsymbol{\theta}_{t})-\varphi(\boldsymbol{\theta}_{t-1})\right)=\sum_{t=1}^T\left(D_\varphi(\boldsymbol{\theta}_t||\boldsymbol{\theta}_{t-1})-\langle\bold{u}_t,\bold{x}_t\rangle\right).$$

Thus, it follows that 
$$\mathcal{R}(\bold{x})-\langle\boldsymbol{\theta}_T,\bold{x}\rangle\geq -\eta\varphi(\bold{0})-\sum_{t=1}^T\left(D_\varphi(\boldsymbol{\theta}_t||\boldsymbol{\theta}_{t-1})-\langle\bold{u}_t,\bold{x}_t\rangle\right).$$

Combining Lemmas \ref{Positive_R_Bound_D} and \ref{Bregman_Bound}, the previous inequality can be rewritten as
$$\sum_{t=1}^T\langle\bold{x}-\bold{x}_t,\bold{u}_t\rangle\leq \mathcal{R}(\bold{x})+\eta\varphi(\bold{0})+\sum_{t=1}^TD_\varphi(\boldsymbol{\theta}_t||\boldsymbol{\theta}_{t-1})\leq \eta\varphi(\bold{0})+{L\over 2\eta}Tu^2_{max}.$$

Because the previous inequality holds for all $\bold{x}\in \Delta_N$ we conclude:
$$\textsc{Regret}_T\leq \eta\varphi(\bold{0})+{L\over2\eta}Tu^2_{max}.$$
\eproof
\subsection{Proof of Theorem \ref{SS_Surplus_algorithm_regret}} The bound (\ref{Bound1}) follows from Lemma \ref{Bound_Regret_FTRL}. To derive Eq. (\ref{Bound2}), define the function $\psi(\eta)=\eta\varphi(\bold{0})+\frac{L}{2\eta}Tu^2_{max}$. Given the strict convexity of $\psi(\eta)$, the first order conditions are necessary and sufficient for  a minimum. In particular, we get $$\psi^\prime(\eta)=\varphi(\bold{0})-\frac{L}{2\eta^2}Tu^2_{max}=0$$
The optimal $\eta$ is given by $\eta^*=\sqrt{{LTu^2_{max}\over2\varphi(\bold{0})}}$. Then, it follows that $\psi(\eta^*)=u_{max}\sqrt{2\varphi(\bold{0})LT}$. Thus  we conclude that

$$\textsc{Regret}_T\leq \psi(\eta^*)=2u_{max}\sqrt{\varphi(\bold{0})LT}.$$\eproof
\subsection{Proof of Lemma \ref{Result_GEV}}This follows from a direct application of  \citet[Thm. 3]{Nesterovetal2019}.\eproof
\subsection{Proof of Theorem \ref{SS_algorithm_GEV}} Combining Lemma \ref{Result_GEV} with Lemma \ref{Bound_Regret_FTRL}
we obtain the bound (\ref{Regret_GEV_SSA}). Following the argument used in proving Theorem \ref{SS_Surplus_algorithm_regret} combined with $L={2M+1\over \eta}$ we obtain the optimized regret bound  (\ref{Regret_GEV_SS}).\eproof

\begin{lemma}\label{GNL_regret_lemma} In the GNL model the following statements hold:
	\begin{itemize}
		\item[i)] The Social Surplus function has a Lipschitz continuous gradient with constant  $({2\over \min_{k}\lambda_k}-1)/\eta.$
		\item[ii)] $\log G(\bold{1})=\log \sum_{k=1}^K\left(\sum_{i=1}^N\alpha^{1/\lambda_k}_{ik} \right)^{\lambda_k}\leq \log N,$
	\end{itemize} 
	
\end{lemma}
\proof i) This follows from a direct application of  \citet[Cor. 4]{Nesterovetal2019}. ii) To prove this, we first show that  for $\lambda_k<\lambda_k^\prime$ we have:

$$\left(\sum_{i=1}^N\alpha^{1/\lambda_k}_{ik} \right)^{\lambda_k}\leq \left(\sum_{i=1}^N\alpha^{1/\lambda^\prime_k}_{ik} \right)^{\lambda^\prime_k}\quad\mbox{for $k=1,\ldots,K.$}$$
Let $p_k={1\over \lambda_k}$ and $p_{k^\prime}={1\over \lambda_{k^\prime}}$, noting that $p^\prime_k<p_k$ whenever $\lambda_k<\lambda_k^\prime$. Using this change of variable, we can write the following ratio

\begin{eqnarray}
{\left(\sum_{i=1}^N\alpha^{p_k}_{ik} \right)^{1/p_k}\over \left(\sum_{j=1}^N\alpha^{p^\prime_k}_{jk} \right)^{1/p^\prime_k}}&=& \left({\sum_{i=1}^N\alpha^{p_k}_{ik} \over \left(\sum_{j=1}^N\alpha^{p^\prime_k}_{jk} \right)^{p_k/p^\prime_k}}\right)^{1/p_k}\nonumber \\
&=& \left(\sum_{i=1}^N\left({\alpha^{p^\prime_k}_{ik} \over \sum_{j=1}^N\alpha^{p^\prime_k}_{jk}}\right)^{p_k/p_k^\prime}\right)^{1/p_k}\nonumber\\
&\leq& \left(\sum_{i=1}^N\left({\alpha^{p^\prime_k}_{ik} \over \sum_{j=1}^N\alpha^{p^\prime_k}_{jk}}\right)\right)^{1/p_k}=1.\nonumber
\end{eqnarray}
The last inequality implies that 
$$\left(\sum_{i=1}^N\alpha^{1/\lambda_k}_{ik} \right)^{\lambda_k}\leq \left(\sum_{i=1}^N\alpha^{1/\lambda^\prime_k}_{ik} \right)^{\lambda^\prime_k}\quad\mbox{for $k=1,\ldots,K.$}$$
Then for $\lambda_k=1$ for $k=1,\ldots, K$, we get
$$\log \sum_{k=1}^K\left(\sum_{i=1}^N\alpha^{1/\lambda_k}_{ik} \right)^{\lambda_k}\leq \log \sum_{k=1}^K\sum_{i=1}^N\alpha_{ik}=\log N.$$
\eproof

\subsection{Proof of Proposition \ref{GNL_regret}} Combining Lemma \ref{GNL_regret_lemma} with Theorem \ref{SS_Surplus_algorithm_regret}, the conclusion follows at once.\eproof

\subsection{Proof of Proposition \ref{R_smooth}}
i) From the definition of $\mathcal{R}(\bold{x})$ we know that $\mathcal{R}(\bold{x})=\eta\varphi^\ast(\bold{x};\eta)$
where $\varphi^\ast(\bold{x};\eta)$ is the convex conjugate of  the parametrized social surplus function $\varphi(\boldsymbol{\theta}/\eta)$. By Lemma \ref{Gradient_Lipschitz}  it follows  that $\varphi(\boldsymbol{\theta}/\eta)$ is $L$-Lipschitz continuous.  Applying  Lemma \ref{Baillon_Haddad} it follows that $\varphi^\ast(\bold{x};\eta)$ is $1/L$ strongly convex. Then it follows that $\mathcal{R}(\bold{x})$ is ${\eta\over L}$-strongly convex. \\
ii) This is a direct implication of Lemma \ref{Convexity_smoothness}.\\
iii) It is easy to see that $ \langle \boldsymbol{\theta},\bold{x}\rangle-\mathcal{R}(\bold{x})$ is a  ${\eta\over L}$-strongly concave function on $\Delta_N$. This implies that  the optimization problem $\max_{\bold{x}\in \Delta_N}\{\langle \boldsymbol{\theta},\bold{x}\rangle-\mathcal{R}(\bold{x})
\}$ must have a unique solution. Let $\bold{x}^*$  be the unique optimal solution. Using the differentiability of $\mathcal{R}(\bold{x})$ combined with the Fenchel equality it follows that $\bold{x}^*=\nabla\varphi(\boldsymbol{\theta})$ iff   $\nabla\varphi(\boldsymbol{\theta})=\arg\max_{\bold{x}\in \Delta_N}\{\langle\boldsymbol{\theta},\bold{x}\rangle-\mathcal{R}(\bold{x})\}$.\eproof
\subsection{Proof of Theorem \ref{No_regret_Discrete_choice}} From Proposition \ref{R_smooth}iii) we know $$\bold{x}_{t+1}=\nabla\varphi(\boldsymbol{\theta}_{t})=\arg\max_{\bold{x}\in\Delta_N}\left\{\langle\boldsymbol{\theta}_t,\bold{x}\rangle-\R(\bold{x})\right\}.$$
 
 This fact implies that  the FRTL algorithm is equivalent to the SSA.  Then the argument used in proving Theorem \ref{SS_Surplus_algorithm_regret} applies. Thus $\textsc{R}^T_{FTRL}$ is bounded by the same term that bounds $\textsc{R}^T_{SSA}$. Similarly, the same optimized bound achieved in $\textsc{R}^T_{SSA}$ applies to $\textsc{R}^T_{FTRL}$. \eproof
 \subsection{Proof of Theorem \ref{Equiv_SSA_FTRL}} From Proposition \ref{R_smooth} we know  $\nabla\varphi(\boldsymbol{\theta}_t)=\bold{x}_{t+1}=\arg\max_{\bold{x}\in \Delta_N}\{\langle\boldsymbol{\theta}_t,\bold{x}\rangle-\R(\bold{x})\}$. Plugging in this observation in the FTRL algorithm the equivalence follows at once.\eproof
 
\subsection{Proof of Proposition \ref{Recursive_Choice_FTRL}} Let us focus at period $t+1$. Accordingly, the associated  Lagrangian is given by:
 	$$\mathcal{L}(\bold{x}_{t+1};\lambda)=\sum_{i=1}^N\boldsymbol{\theta}_{it} \bold{x}_{it+1}-{\eta}\sum_{i=1}^N\bold{x}_{it+1}\log\Phi_i(\bold{x}_{t+1})+\lambda\left(\sum_{i=1}^N\bold{x}_{it+1}-1\right).$$
 	From \cite[Prop. A1ii)]{Fosgerauetal2019a} we know that $\Phi(\bold{x}_{t+1})$ is differentiable with 
 	$$\sum_{j=1}^N\bold{x}_{jt+1}{\partial \Phi_j(\bold{x}_{t+1})\over \partial \bold{x}_{it+1}}=1\quad\forall i\in A.$$
 	Using this fact, the set of  first order conditions can be written as:
 	\begin{eqnarray}
 		{\partial \mathcal{L}(\bold{x}_{t+1};\lambda)\over \partial\bold{x}_{t+1}}&=&\boldsymbol{\theta}_t-{ \eta}\log \Phi(\bold{x}_{t+1})-{ \eta}+\lambda=0.\label{FOC11}\\
 		{\partial \mathcal{L}(\bold{x}_{t+1};\lambda)\over \partial\lambda}&=&\sum_{i=1}^n\bold{x}_{it+1}-1=0.\label{FOC22}
 	\end{eqnarray}
 	Noting that  $\boldsymbol{\theta}_t=\bold{u}_t+\boldsymbol{\theta}_{t-1}$,  Eq. (\ref{FOC11}) can be expressed as
 	\begin{eqnarray}\label{FOC33}
 		e^{\bold{u}_t/\eta+\boldsymbol{\theta}_{t-1}/\eta}e^{\lambda/\eta-1}&=&\Phi(\bold{x}_{t+1})
 	\end{eqnarray}
 	Recalling that $\Phi(\cdot)=H^{-1}(\cdot)$, from (\ref{FOC33}) we get:
 	\begin{eqnarray}
 		H(e^{\bold{u}_t/\eta+\boldsymbol{\theta}_{t-1}/\eta}e^{\lambda/\eta-1})&=&\bold{x}_{t+1}.\nonumber
 	\end{eqnarray}
 	Noting that $H(\cdot)$ is homogeneous of degree 1, we get:
 	\begin{eqnarray}
 		H(e^{\bold{u}_t/\eta+\boldsymbol{\theta}_{t-1}/\eta})e^{\lambda/\eta-1}&=&\bold{x}_{t+1}.\nonumber
 	\end{eqnarray}
 	Using (\ref{FOC22}) we find that 
 	\begin{eqnarray}
 		e^{\lambda/\eta-1}&=&{1\over \sum_{j=1}^NH_j(e^{\bold{u}_t/\eta+\boldsymbol{\theta}_{t-1}/\eta})}.\nonumber
 	\end{eqnarray}
 	Then it is easy to see that 
 	\begin{equation}\label{FOC44}
 		\bold{x}_{it+1}={H_i(e^{\bold{u}_t/\eta+\boldsymbol{\theta}_{t-1}/\eta})\over \sum_{j=1}^NH_j(e^{\bold{u}_t/\eta+\boldsymbol{\theta}_{t-1}/\eta})}\quad \mbox{for all  $i\in A, t\geq 1.$}
 	\end{equation}
 	From (\ref{FOC44}), it is easy to see that at period $t$ we must have:
 	\begin{equation}\label{FOC55}
 		\bold{x}_{it}={H_i(e^{\boldsymbol{\theta}_{t-1}/\eta})\over \sum_{j=1}^NH_j(e^{\boldsymbol{\theta}_{t-1}/\eta})}\quad \mbox{for all  $i\in A$.}
 	\end{equation}
 	Once again, using the fact that $\Phi(\cdot)$ is homogeneous of degree 1,  in (\ref{FOC55}) we find:
 	\begin{equation}\label{FOC51}
 		\Phi(\bold{x}_{t})\sum_{j=1}^NH_j(e^{\boldsymbol{\theta}_{t-1}/
 			\eta})=e^{\boldsymbol{\theta}_{t-1}/\eta}\quad \mbox{for  all $i\in A$.}
 	\end{equation}
 	Define $w_{t-1}\triangleq\log\left(\sum_{j=1}^NH_j(e^{\boldsymbol{\theta}_{t-1}/\eta})\right)$. Using this definition, combined with Eq. (\ref{FOC44}),  we get:
 	\begin{equation}
 		\bold{x}_{it+1}={H_i(\Phi(\bold{x}_{t})e^{\bold{u}_t/\eta+ w_{t-1}})\over\sum_{j=1}^N H_j(\Phi(\bold{x}_{t})e^{\bold{u}_t/\eta+w_{t-1}})}.\nonumber
 	\end{equation}
 	Finally, using the homogeneity of $H$, we conclude that 
 	\begin{equation}
 		\bold{x}_{it+1}={H_i(e^{\bold{u}_t/\eta+\alpha(\bold{x}_t)})\over\sum_{j=1}^N H_j(e^{\bold{u}_t/\eta+\alpha(\bold{x}_t)})}, \quad\forall i\in A, t\geq 1.\nonumber
 	\end{equation}
 	\eproof 
\subsection{Proof of Lemma \ref{NL_regul_lemma}} This follows from \citet{Fosgerauetal2019a}.\eproof


\vspace{2ex}

In proving the Proposition \ref{Exp_Nested1} we make use of the following technical lemma.
\begin{lemma}\label{Inverse_phi} Consider the NL model. Define the vector valued function $\Phi:\Delta_N\longrightarrow\RR_+^N$ where the $i$-th component  is defined as:
	\begin{equation}\label{phi_formula}
	\Phi_i(\bold{x})=\bold{x}_i^{\lambda_k}\left(\sum_{j\in\mathcal{N}_k}\bold{x}_j\right)^{1-\lambda_k}\quad \mbox{for all $i\in\mathcal{N}_k,k=1,\ldots,K$},
	\end{equation}
	Then $\Phi(\bold{x})$ is invertible and homogeneous of degree 1.		
\end{lemma}

\proof This follows from  \citet[Prop. 8]{Fosgerauetal2019a}.  \eproof
\subsection{Proof of Proposition \ref{Exp_Nested1}} Noting that $\mathcal{R}(\bold{x}_{t+1})=\langle\bold{x}_{t+1},\log\Phi(\bold{x}_{t+1})\rangle$, we can write the associated  Lagrangian as:
$$\mathcal{L}(\bold{x}_{t+1};\boldsymbol{\theta}_t,\mu,\eta)=\langle\boldsymbol{\theta}_t, \bold{x}_{t+1}\rangle-{ \eta}\langle\bold{x}_{t+1},\log\Phi(\bold{x}_{t+1})\rangle+\mu\left(\sum_{i=1}^n\bold{x}_{it+1}-1\right).$$
Given that $\mathcal{R}(\bold{x})$ is ${1\over L}$-strongly convex, first order conditions are necessary and sufficient for the existence and uniqueness of a maximum. In particular, we find the maximizer $\bold{x}_{t+1}$ by solving:
\begin{eqnarray}
{\partial \mathcal{L}(\bold{x}_{t+1};\boldsymbol{\theta}_t,\mu,\eta)\over \partial\bold{x}_{t+1}}&=&\boldsymbol{\theta}_t-{\eta}\log \Phi(\bold{x}_{t+1})-{ \eta}+\mu=0.\label{FOC1}\nonumber\\
{\partial \mathcal{L}(\bold{x}_{t+1};\boldsymbol{\theta}_t,\mu,\eta)\over \partial\lambda}&=&\sum_{i=1}^n\bold{x}_{it+1}-1=0.\label{FOC2}\nonumber
\end{eqnarray}
From the definition of the cumulative payoff vector,  it follows that $\boldsymbol{\theta}_t=\bold{u}_t+\boldsymbol{\theta}_{t-1}$. Using this fact we get
\begin{eqnarray}\label{FOC3}
e^{\bold{u}_t/\eta+\boldsymbol{\theta}_{t-1}/\eta}e^{\mu/\eta-1}&=&\Phi(\bold{x}_{t+1}).\nonumber
\end{eqnarray}
Using Lemma \ref{Inverse_phi} we obtain:
\begin{eqnarray}
H(e^{\bold{u}_t/\eta+\boldsymbol{\theta}_{t-1}/\eta}e^{\mu/\eta-1})&=&\bold{x}_{t+1},\nonumber
\end{eqnarray}
where $H(\cdot)\triangleq \Phi^{-1}(\cdot).$
Noting that $H(\cdot)$ is homogeneous of degree 1, the previous expression can be rewritten as:
\begin{eqnarray}
H(e^{\bold{u}_t/\eta+\boldsymbol{\theta}_{t-1}/\eta})e^{\mu/\eta-1}&=&\bold{x}_{t+1}.\nonumber
\end{eqnarray}
Now using the constraint $\sum_{i=1}^N\bold{x}_{it+1}=1$ we get
\begin{eqnarray}
e^{\mu/\eta-1}&={1\over \sum_{k=1}^K\sum_{j\in\mathcal{N}_k}H_j(e^{\bold{u}_t/\eta+\boldsymbol{\theta}_{t-1}/\eta})}&.\nonumber
\end{eqnarray}
Then we find that for all $t$ 
\begin{equation}\label{FOC42}
\bold{x}_{it+1}={H_i(e^{\bold{u}_t/\eta+\boldsymbol{\theta}_{t-1}/\eta})\over \sum_{k=1}\sum_{j\in\mathcal{N}_k}H_j(e^{\bold{u}_t/\eta+\boldsymbol{\theta}_{t-1}/\eta})}\quad \mbox{for  $i\in\mathcal{N}_k,k=1,\ldots,K$.}
\end{equation}
In the previous expression, we note that 
\begin{eqnarray}
H_i(e^{\bold{u}_t/\eta+\boldsymbol{\theta}_{t-1}/\eta})&=&\left(\sum_{j\in\mathcal{N}_k}e^{(\bold{u}_{jt}+\boldsymbol{\theta}_{jt-1})/\eta\lambda_k}\right)^{\lambda_k-1}e^{(\bold{u}_{it}+\boldsymbol{\theta}_{it-1})/\eta\lambda_k},\nonumber\\
\sum_{k=1}^K\sum_{j\in \mathcal{N}_k}H_j(e^{\bold{u}_t/\eta+\boldsymbol{\theta}_{t-1}/\eta})&=&\sum_{k=1}^K\left(\sum_{j\in\mathcal{N}_k}e^{(\bold{u}_{jt}+\boldsymbol{\theta}_{jt-1})/\eta\lambda_k}\right)^{\lambda_k}.
\nonumber
\end{eqnarray}

From (\ref{FOC42}) it is easy to see that for period $t$ we must have:

\begin{equation*}\label{FOC5}
	\bold{x}_{it}={H_i(e^{\boldsymbol{\theta}_{t-1}/\eta})\over \sum_{k=1}^K\sum_{j\in\mathcal{N}_k}H_j(e^{\boldsymbol{\theta}_{t-1}/\eta})}\quad \mbox{for  $i\in\mathcal{N}_k,k=1,\ldots,K.$}
\end{equation*}

Using once again using the fact that $\Phi(\cdot)$ is homogenous of degree 1, we get 
\begin{equation}\label{FOC5}
\Phi(\bold{x}_{t})\sum_{k=1}^K\sum_{j\in\mathcal{N}_k}H_j(e^{\boldsymbol{\theta}_{t-1}/\eta})=e^{\boldsymbol{
		\theta}_{t-1}/\eta}\quad  \mbox{for  $i\in\mathcal{N}_k,k=1,\ldots,K.$}\nonumber
\end{equation}

Taking log on both sides in the previous expression we obtain
\begin{equation}\label{FOC6}
\log\Phi(\bold{x}_{t})+\log\left(\sum_{k=1}^K\sum_{j\in\mathcal{N}_k}H_j(e^{\boldsymbol{\theta}_{t-1}/\eta})\right)=\boldsymbol{\theta}_{t-1}/
\eta\quad\mbox{for  $i\in\mathcal{N}_k,k=1,\ldots,K.$}\nonumber
\end{equation}

Define $m_{t-1}\triangleq\log\left(\sum_{k=1}^K\sum_{j\in\mathcal{N}_k}H_j(e^{\boldsymbol{
		\theta}_{t-1}/\eta})\right)$. Using this definition we get:
\begin{equation}
\bold{x}_{it+1}={H_i(\Phi(\bold{x}_{t})e^{\bold{u}_t/\eta+ m_{t-1}})\over\sum_{k=1}^K\sum_{j\in\mathcal{N}_k} H_j(\Phi(\bold{x}_{t})e^{\bold{u}_t/\eta+m_{t-1}})}.\nonumber
\end{equation}

Using the homogeneity of $H$ one last time, we obtain: 
\begin{equation}
\bold{x}_{it+1}={H_i(\Phi_i(\bold{x}_t)e^{\bold{u}_t/
		\eta})\over\sum_{k=1}^K\sum_{j\in\mathcal{N}_k} H_j(\Phi_j(\bold{x}_t)e^{\bold{u}_t/\eta})}\nonumber \quad\mbox{for  $i\in\mathcal{N}_k,k=1,\ldots,K.$}\nonumber
\end{equation}

Finally, replacing the expression for $H_i$ using the NL assumption, the conclusion follows at once. \eproof 
\subsection{Proof of Corollary} When $\lambda_{k}=1$ for all $k\in K$, we know that the NL boils to the MNL model. Thus the conclusion follows at once.\eproof
\vspace{2ex}
\newpage

\section{Online material not for publication}
\begin{lemma}\label{Bound_OFTRL}Let Assumptions \ref{Shocks_Assumption} and \ref{Gradient_LL} hold.  Then in the OFTRL algorithm the following hold:
	\begin{equation}\label{Regret_OFTRL}
	\sum_{t=1}^T\langle\bold{x}^*-\bold{x}_t,\bold{u}_t\rangle\leq \eta\varphi(\bold{0})+{L\over 2\eta}\sum_{t=1}^{T}\left\|\mathbf{u}_{t}-\boldsymbol{\beta}_{t}\right\|_{*}^{2}.
	\end{equation}
\end{lemma}
\proof The proof of this Lemma follows from a simple adaptation of \cite[Lemma 2]{RakhlinSridharan2013}.\eproof

\begin{lemma}\label{Cases_regret}In the OFTRL the following statements hold
	\begin{itemize}
		\item[i)] In the $S$-step recency bias:
		$$\sum_{t=1}^T\|\bold{u}_t-\boldsymbol{\beta}_t\|^2_\ast\leq S^2\sum_{t=1}^T\|\bold{u}_t-\bold{u}_t\|^2_\ast$$
		\item[ii)] In the geometrically discounted recency bias we have:
		
		$$\sum_{t=1}^T\|\bold{u}_t-\boldsymbol{\beta}_t\|^2_\ast\leq 	\frac{1}{(1-\delta)^{3}} \sum_{t=1}^{T}\left\|\mathbf{u}_{t}-\mathbf{u}_{t-1}\right\|_{*}^{2}$$
	\end{itemize}
	
\end{lemma}

\proof In proving parts i)  and ii) we follow the proof of Lemma 21 in \citet{NIPS2015_5763}. Concretely, in proving part i) we have the following:

\begin{eqnarray}
\sum_{t=1}^T\|\bold{u}_t-\boldsymbol{\beta}_t\|^2_\ast&=& \sum_{t=1}^{T}\left\|\mathbf{u}_{t}-\frac{1}{H} \sum_{\tau=t-H}^{t-1} \mathbf{u}_{\tau}\right\|_{*}^{2},\nonumber\\
&=&  \sum_{t=1}^{T}\left(\frac{1}{H} \sum_{\tau=t-H}^{t-1}\left\|\mathbf{u}_{t}-\mathbf{u}_{\tau}\right\|_{*}\right)^{2}.\label{Eq.S.Steps}
\end{eqnarray}

By the triangle inequality we get
\begin{eqnarray}
\frac{1}{H} \sum_{\tau=t-H}^{t-1}\left\|\mathbf{u}_{t}-\mathbf{u}_{\tau}\right\|_{\ast}&\leq & \frac{1}{H} \sum_{\tau=t-H}^{t-1} \sum_{q=\tau}^{t-1}\left\|\mathbf{u}_{q+1}-\mathbf{u}_{q}\right\|_{*}\nonumber\\
&=& \sum_{\tau=t-H}^{t-1} \frac{t-\tau}{H}\left\|\mathbf{u}_{\tau+1}-\mathbf{u}_{\tau}\right\|_{*} \leq \sum_{\tau=t-H}^{t-1}\left\|\mathbf{u}_{\tau+1}-\mathbf{u}_{\tau}\right\|_{*}\nonumber
\end{eqnarray}

By the Cauchy-Schwarz inequality, we get:
$$
\left(\sum_{\tau=t-H}^{t-1}\left\|\mathbf{u}_{\tau+1}-\mathbf{u}_{\tau}\right\|_{*}\right)^{2} \leq H \sum_{\tau=t-H}^{t-1}\left\|\mathbf{u}_{\tau+1}-\mathbf{u}_{\tau}\right\|_{*}^{2}.
$$
Thus it follows that:
\begin{eqnarray}
\sum_{t=1}^T\|\bold{u}_t-\boldsymbol{\beta}_t\|^2_\ast&\leq& H \sum_{t=1}^{T} \sum_{\tau=t-H}^{t-1}\left\|\mathbf{u}_{\tau+1}-\mathbf{u}_{\tau}\right\|_{*}^{2}\nonumber\\
&\leq& H^{2} \sum_{t=1}^{T}\left\|\mathbf{u}_{t}-\mathbf{u}_{t-1}\right\|_{*}^{2}\nonumber
\end{eqnarray}
In proving part (ii) we follow the proof of Lemma 22 in \citet{NIPS2015_5763}. We begin noting that
\begin{eqnarray}
\sum_{t=1}^{T}\left\|\mathbf{u}_{t}-\boldsymbol{\beta}_{t}\right\|_{*}^{2}&=& \sum_{t=1}^{T}\left\|\mathbf{u}_{t}-\frac{1}{\sum_{\tau=0}^{t-1} \delta^{-\tau}} \sum_{\tau=0}^{t-1} \delta^{-\tau} \mathbf{u}_{\tau}\right\|_{\ast}^{2}\nonumber
\end{eqnarray}
We want to show that 
$$\sum_{t=1}^{T}\left\|\mathbf{u}_{t}-\frac{1}{\sum_{\tau=0}^{t-1} \delta^{-\tau}} \sum_{\tau=0}^{t-1} \delta^{-\tau} \mathbf{u}_{\tau}\right\|_{*}^{2} \leq \frac{1}{(1-\delta)^{3}} \sum_{t=1}^{T}\left\|\mathbf{u}_{t}-\mathbf{u}_{t-1}\right\|_{*}^{2}$$
In proving this, we first note that
\begin{eqnarray}
\left\|\mathbf{u}_{t}-\frac{1}{\sum_{\tau=0}^{t-1} \delta^{-\tau}} \sum_{\tau=0}^{t-1} \delta^{-\tau} \mathbf{u}_{\tau}\right\|&=& \frac{1}{\sum_{\tau=0}^{t-1} \delta^{-\tau}} \sum_{\tau=0}^{t-1} \delta^{-\tau}\left\|\mathbf{u}_{t}-\mathbf{u}_{\tau}\right\|_{*}\nonumber\\
&\leq & \frac{1}{\sum_{\tau=0}^{t-1} \delta^{-\tau}} \sum_{\tau=0}^{t-1} \delta^{-\tau} \sum_{q=\tau}^{t-1}\left\|\mathbf{u}_{q+1}-\mathbf{u}_{q}\right\|_{*}\nonumber\\
&=& \frac{1}{\sum_{\tau=0}^{t-1} \delta^{-\tau}} \sum_{a=0}^{t-1}\left\|\mathbf{u}_{q+1}-\mathbf{u}_{q}\right\|_{*} \sum_{\tau=0}^{q} \delta^{-\tau}\nonumber\\
&=& \frac{1}{\sum_{\tau=0}^{t-1} \delta^{-\tau}} \sum_{q=0}^{t-1}\left\|\mathbf{u}_{q+1}-\mathbf{u}_{q}\right\|_{*} \delta^{-q} \frac{1-\delta^{q+1}}{1-\delta}\nonumber\\
&\leq& \frac{1}{1-\delta} \frac{1}{\sum_{\tau=0}^{t-1} \delta^{-\tau}} \sum_{q=0}^{t-1} \delta^{-q}\left\|\mathbf{u}_{q+1}-\mathbf{u}_{q}\right\|_{*}\nonumber
\end{eqnarray}

Second, by Cauchy-Schwartz  we have

\begin{eqnarray}
\left(\frac{1}{1-\delta} \frac{1}{\sum_{\tau=0}^{t-1} \delta^{-\tau}} \sum_{q=0}^{t-1} \delta^{-q}\left\|\mathbf{u}_{q+1}-\mathbf{u}_{q}\right\|_{*}\right)^{2}&=& \frac{1}{(1-\delta)^{2}} \frac{1}{\left(\sum_{\tau=0}^{t-1} \delta^{-\tau}\right)^{2}}\left(\sum_{q=0}^{t-1} \delta^{-q / 2} \cdot \delta^{-q / 2}\left\|\mathbf{u}_{q+1}-\mathbf{u}_{q}\right\|_{*}\right)^{2}\nonumber\\
&\leq& \frac{1}{(1-\delta)^{2}} \frac{1}{\left(\sum_{\tau=0}^{t-1} \delta^{-\tau}\right)^{2}} \sum_{q=0}^{t-1} \delta^{-q} \cdot \sum_{q=0}^{t-1} \delta^{-q}\left\|\mathbf{u}_{q+1}-\mathbf{u}_{q}\right\|_{*}^{2}\nonumber\\
&=& \frac{1}{(1-\delta)^{2}} \frac{1}{\sum_{\tau=0}^{t-1} \delta^{-\tau}} \sum_{q=0}^{t-1} \delta^{-q}\left\|\mathbf{u}_{q+1}-\mathbf{u}_{q}\right\|_{*}^{2}\nonumber\\
&= & \frac{1}{(1-\delta)^{2}} \frac{1}{\sum_{\tau=0}^{t-1} \delta^{t-\tau}} \sum_{q=0}^{t-1} \delta^{t-q}\left\|\mathbf{u}_{q+1}-\mathbf{u}_{q}\right\|_{*}^{2}\nonumber\\
&\leq & \frac{1}{\delta(1-\delta)^{2}} \sum_{q=0}^{t-1} \delta^{t-q}\left\|\mathbf{u}_{q+1}-\mathbf{u}_{q}\right\|_{*}^{2}\nonumber
\end{eqnarray}
Combining previous expressions we get:
$$
\left\|\mathbf{u}_{t}-\frac{1}{\sum_{\tau=0}^{t-1} \delta^{-\tau}} \sum_{\tau=0}^{t-1} \delta^{-\tau} \mathbf{u}_{\tau}\right\|_{*}^{2} \leq \frac{1}{\delta(1-\delta)^{2}} \sum_{q=0}^{t-1} \delta^{t-q}\left\|\mathbf{u}_{q+1}-\mathbf{u}_{q}\right\|_{*}^{2}
$$
Summing over all  $t$ and re-arranging we get: 

\begin{eqnarray}
\sum_{t=1}^{T}\left\|\mathbf{u}_{t}-\frac{1}{\sum_{\tau=0}^{t-1} \delta^{-\tau}} \sum_{\tau=0}^{t-1} \delta^{-\tau} \mathbf{u}_{\tau}\right\|^{2}&\leq& \frac{1}{\delta(1-\delta)^{2}} \sum_{t=1}^{T} \sum_{q=0}^{t-1} \delta^{t-q}\left\|\mathbf{u}_{q+1}-\mathbf{u}_{q}\right\|_{*}^{2}\nonumber \\
&=& \frac{1}{\delta(1-\delta)^{2}} \sum_{q=0}^{T-1} \delta^{-q}\left\|\mathbf{u}_{q+1}-\mathbf{u}_{q}\right\|_{*}^{2} \sum_{t=q+1}^{T} \delta^{t}\nonumber\\
&=& \frac{1}{\delta(1-\delta)^{2}} \sum_{q=0}^{T-1} \delta^{-q}\left\|\mathbf{u}_{q+1}-\mathbf{u}_{q}\right\|_{*}^{2} \frac{\delta\left(\delta^{q}-\delta^{T}\right)}{1-\delta}\nonumber\\
&=& \frac{1}{(1-\delta)^{3}} \sum_{q=0}^{T-1}\left\|\mathbf{u}_{q+1}-\mathbf{u}_{q}\right\|_{*}^{2}\left(1-\delta^{T-q}\right)\nonumber\\
&\leq& \frac{1}{(1-\delta)^{3}} \sum_{q=0}^{T-1}\left\|\mathbf{u}_{q+1}-\mathbf{u}_{q}\right\|_{*}^{2}\nonumber 
\end{eqnarray}
\eproof
\subsection{Proof of Proposition \ref{OFTRL_Regret}} i) Combining Lemma \ref{Bound_OFTRL} and Assumption (\ref{Bound_Dual})  we know that 
$$\textsc{R}_{OFTRL}^T\leq \eta\varphi(\bold{0})+\frac{L}{2\eta}TB^2.$$
Optimizing over $\eta$ we find that the regret is minimized at  $\eta=\sqrt{\frac{LTB^2}{2\varphi(\bold{0})}}$.  Thus we obtain
$$\textsc{R}_{OFTRL}^T\leq B\sqrt{2LT\varphi(\bold{0})}. $$
ii) Combining Lemma \ref{Bound_OFTRL} and Assumption (\ref{Bound_Dual})  we know that 
$$\textsc{R}_{OFTRL}^T\leq \eta\varphi(\bold{0})+{L\over 2\eta}\sum_{t=1}^{T}\left\|\mathbf{u}_{t}-\boldsymbol{\beta}_{t}\right\|_{*}^{2}.$$  
Using Lemma \ref{Cases_regret}i) combined with Eq. (\ref{Bound_Dual})  we obtain that:
$$ \textsc{R}_{OFTRL}^T\leq \eta\varphi(\bold{0})+\frac{L}{2\eta}TS^2B^2$$
Optimizing over $\eta$ we find that the regret is minimized at   $\eta=\sqrt{\frac{LTS^2B^2}{2\varphi(\bold{0})}}$. In particular, we get:
$$\textsc{R}_{OFTRL}^T\leq SB\sqrt{2LT\varphi(\bold{0})}.$$
In showing part iii), we note that combining Lemma \ref{Bound_OFTRL},  Lemma \ref{Cases_regret}ii), and condition (\ref{Bound_Dual})  
$$ \textsc{R}_{OFTRL}^T\leq \eta\varphi(\bold{0})+\frac{L}{2\eta(1-\delta)^3}TB^2$$
Optimizing over $\eta$ we find that the regret is minimized at   $\eta=\sqrt{\frac{LTB^2}{2(1-\delta)^3\varphi(\bold{0})}}$. Thus, we get:
$$\textsc{R}_{OFTRL}^T\leq B\sqrt{\frac{2LT\varphi(\bold{0})}{(1-\delta)^3}}.$$\eproof

\vspace{2ex}
\subsection{Proof of Proposition \ref{SS_equiv_CBM}} We proof this result using Theorem \ref{ChenWortmanThm}. Differentiability follows from the Williams-Daly-Zachary theorem. Now let $\bold{q}$ and $\bold{q}^\prime$ with $\bold{q}^\prime\geq \bold{q}$. Note that   $\max_{i=1m\ldots,N}\{\bold{q}^\prime_i+\epsilon_i\}\geq \max_{i=1m\ldots,N}\{\bold{q}_i+\epsilon_i\} $. Taking expectation with respect to $\epsilon$, it follows that $\varphi(\bold{q}^\prime)\geq \varphi(\bold{q})$, which implies the increasing monotonicity of $\varphi$. Finally, positive translation invariance follows from the fact that 
$\varphi(\bold{q}+k\bold{1})=\EE(\max_{i=1,\ldots,N}\{\bold{q}_i+k+\epsilon_i\})=\EE(\max_{i=1,\ldots,N}\{\bold{q}_i+\epsilon_i\})+k=\varphi(\bold{q})+k$.\eproof
\newpage
\subsection{Applications of the GNL model}\label{AppendixGNL}
\subsubsection{The Cross Nested Logit (CNL) model} \citet{Vosha1997} introduces the
cross nested logit model. The main assumption of this model is that $\lambda_k=\lambda$ for all $k=1,\ldots,K$. Thus, the generator $G$ boils down to the expression: $$G(e^{\boldsymbol{\theta}_t/\eta})=\sum_{k=1}^K\left(\sum_{i=1}^N\left(\alpha_{ik}e^{\boldsymbol{\theta}_{it}/\eta}\right)^{1/\lambda}\right)^\lambda.$$
In addition,  in this case the  constant $M$ is given by $M={2\over \lambda}-1$. Accordingly, the Social Surplus function has Lipschitz continuous gradient with constant $({2\over \lambda}-1)/\eta.$
\subsubsection{The Paired Combinatorial Logit (PCL) model} In this model each pair of alternatives is represented by a nest. Formally, the set of nests is defined as $\mathcal{N}=\left\{(i,j)\in A\times A: i\neq j\right \}$.  Accordingly, we  define 
$$\alpha_{ik}=\begin{cases} {1\over 2(N-1)} &\mbox{if $k=(i, j),(j, i)$ with $j \neq i$ }  \\ 
0& \mbox{otherwise}  \end{cases} .
$$

Using the previous expression, the  generator $G$ can be written as:
$$G(e^{\boldsymbol{\theta}_t/\eta})=\sum_{k=(i,j)\in \mathcal{N}}\left((\alpha_{ik}e^{\boldsymbol{\theta}_{it}/\eta})^{1/\lambda_k}+(\alpha_{jk}e^{\boldsymbol{\theta}_{jt}/\eta})^{1/\lambda_k}\right)^{\lambda_k}.$$

In this case   the Lipschitz constant is  $({2\over \min\lambda_k}-1)/\eta.$

 In addition, $\log G(\bold{1})\leq\log N$. Then the regret analysis follows from Proposition \ref{GNL_regret}.
\subsubsection{The Ordered GEV (OGEV) model} \citet{Small1987} studies a GEV model in which the alternatives  are allocated to nests based on their \emph{proximity} in an ordered set.  Following \citet{Small1987} we define the set of overlapping nests to be
$$\mathcal{N}=\{1,\ldots,N+N^\prime\},$$
with $\alpha_{i\ell}>0$ for all $\ell\in\{i,\ldots,N+N^\prime\}$ and $\alpha_{i\ell}=0$ for $i\in\mathcal{N}\setminus \{i,\ldots,N+N^\prime\}$, and each alternative lies exactly in $N^\prime+1$ of these nests. 
In the model there are $N+N^\prime$ overlapping nests. Each nest $\ell\in \mathcal{N}$ is defined as $\mathcal{N}_\ell=\{i\in A:l-m\leq i\leq \ell \}$
where $i\in\mathcal{N}_l$ for $\ell=i,\ldots,i+N^\prime$. Despite this rather complex description, the generator function $G$ takes the familiar form:
$$G(e^{\boldsymbol{\theta}_t/\eta})=\sum_{k=1}^{N+N^\prime}\left(\sum_{i\in \mathcal{N}_k}(\alpha_{ik}e^{\boldsymbol{\theta}_{it}/\eta})^{	1/\lambda_k}\right)^{\lambda_k}$$
Thus in this case  the Lipschitz constant is given by $({2\over \min_{k=1,\ldots,K}\lambda_k}-1)/\eta$. Moreover,  $\log G(\bold{1})\leq\log N$. Thus Proposition \ref{GNL_regret} applies and we conclude that the  OGEV model is Hannan consistent.
\subsubsection{ Principles of  Differentiation GEV model (PDGEV)}\citet{Bresnahanetal1997} introduce  the PDGEV model. This appproach is based on the idea of  markets for differentiated products. Using this idea, the set  of nests is defined in terms of the attributes that characterize the different products (goods). For instance, in  the context of transportation modeling, the attributes can include  mode to work, destination, number of cars, and residential location. Accordingly, let $D$ be the set of attributes with $\mathcal{N}=\bigcup_{d\in D}\mathcal{N}_d$  and $\mathcal{N}_d=\{k\in \mathcal{N}:\mbox{nest $k$  contains products with attribute $d$}\}$ be the nest that contains the alternatives with attribute $d.$ Similarly, let $\mathcal{N}_{kd}$ denote the nest $k$ with attribute $d$.
$$\alpha_{ik}=\left\{\begin{array}{cl}\alpha_{d} & \text { if } i \in N_{kd} \text { and } k \in \mathcal{N}_{d} \\ 0 & \text { otherwise}\end{array}\right.$$
\smallskip

In this case the  generator $G$ takes the form:
$$G(e^{\boldsymbol{\theta}_t/\eta})=\sum_{d\in D}\alpha_d\sum_{k\in \mathcal{N}_d}\left(\sum_{i\in \mathcal{N}_{kd}}e^{\boldsymbol{\theta}_{it}/\eta\lambda_d}   \right)^{\lambda_d}.$$

It is easy to see that the previous generator is a particular case of the GNL model. Furthermore, the Lipschitz constant is  $({2\over min_{d=1,\ldots,D}\lambda_d}-1)/\eta$ and $\log G(\bold{1})\leq\log N$. Thus  Proposition \ref{GNL_regret} applies in a direct way.
\smallskip

We close this appendix summarizing the regret bounds  for  the models  described  in the main text and  in this appendix. Table \ref{Summary} below makes explicit our regret analysis. The table displays how several GEV models shares  the same optimized regret bound.
\begin{table}[h!]
	\begin{center}
		
		\begin{tabular}{l|c|r} 
			\textbf{Model} & \textbf{Optimal $\eta$} & \textbf{Regret Bound}\\
			\hline
			RUM & $\sqrt{{LTu_{max}^2\over 2\varphi(0)}}$ & $u_{max}\sqrt{2\varphi(0)LT}$\\
			GEV & $\sqrt{{(2M+1)Tu_{max}^2\over 2\log G(1)}}$ & $u_{max}\sqrt{2 \log G(\textbf{1})(2M+1)T}$\\
			GNL & $\sqrt{\left( {2\over \min_{k}\lambda_k}-1\right)Tu^2_{max}\over 2\log N}$ & $u_{max}\sqrt{2\log N \left( {2\over \min_{k}\lambda_k}-1\right)T}$\\
			PCL&$\sqrt{\left( {2\over \min_{k}\lambda_k}-1\right)Tu^2_{max}\over 2\log N}$  & $u_{max}\sqrt{2\log N \left( {2\over \min_{k}\lambda_k}-1\right)T}$\\
			CNL&$\sqrt{\left( {2\over \lambda}-1\right)Tu^2_{max}\over 2\log N}$ &  $u_{max}\sqrt{2\log N \left( {2\over \lambda}-1\right)T}$\\
			
			OGEV&$\sqrt{\left( {2\over \min_{k}\lambda_k}-1\right)Tu^2_{max}\over 2\log N}$ & $u_{max}\sqrt{2\log N \left( {2\over \min_{k}\lambda_k}-1\right)T}$\\
			PDGEV&$\sqrt{\left( {2\over \min_{d}\lambda_d}-1\right)Tu^2_{max}\over 2\log N}$ & $u_{max}\sqrt{2\log N \left( {2\over \min_{d}\lambda_d}-1\right)T}$\\
			NL& $\sqrt{\left( {2\over \min_{k}\lambda_k}-1\right)Tu^2_{max}\over 2\log N}$ & $u_{max}\sqrt{2\log N \left( {2\over \min_{k}\lambda_k}-1\right)T}$\\
			Logit&$\sqrt{Tu^2_{max}\over 2\log N}$ & $u_{max}\sqrt{2\log N T}$

		\end{tabular}
		\vspace{4ex}
		\caption{Summary of the optimized regret bound for the RUM and several GEV models.}
		\label{Summary}
		
	\end{center}
\end{table}


\end{document}